\documentclass[fleqn,usenatbib]{mnras} 
\usepackage{txfonts}

\usepackage[T1]{fontenc}
\usepackage{ae,aecompl}

\usepackage{color, soul, ulem,graphicx, amsbsy}

\title[Evolution of atmospheric escape in close-in giants]{Evolution of atmospheric escape in close-in giant planets and their associated Ly$\alpha$ and H$\alpha$ transit predictions}

\author[Allan and Vidotto]{
A.~Allan\thanks{E-mail: allana@tcd.ie}  and A.~A.~Vidotto \\
School of Physics, Trinity College Dublin, the University of Dublin, College Green, Dublin-2, Ireland
}

\date{Accepted XXX. Received YYY; in original form ZZZ}

\pubyear{2019}

\begin{document}
\label{firstpage}
\pagerange{\pageref{firstpage}--\pageref{lastpage}}
\maketitle

\begin{abstract}
Strong atmospheric escape has been detected in several close-in exoplanets. As these planets consist mostly of hydrogen, observations in hydrogen lines, such as Ly$\alpha$ and H$\alpha$, are powerful diagnostics of escape. Here, we simulate the evolution of atmospheric escape of close-in giant planets and calculate their associated  Ly$\alpha$ and H$\alpha$ transits. We use a one-dimensional hydrodynamic escape model to compute physical properties of the atmosphere and a ray-tracing technique to simulate spectroscopic transits. We consider giant (0.3 and 1$M_{\rm jup}$) planets orbiting a solar-like star at 0.045au, evolving from 10 to 5000 Myr. We find that younger giants show higher rates of escape, owing to a favourable combination of higher irradiation fluxes and weaker gravities. Less massive planets show higher escape rates ($10^{10}$ -- $10^{13}$ g/s) than those more massive ($10^9$ -- $10^{12}$ g/s) over their evolution. We estimate that the 1-$M_{\rm jup}$ planet would lose at most 1\% of its initial mass due to escape, while the 0.3-$M_{\rm jup}$ planet, could lose up to 20\%. This supports the idea that the Neptunian desert has been formed due to significant mass loss in low-gravity planets.  At younger ages, we find that the mid-transit Ly$\alpha$ line is saturated at line centre, while H$\alpha$ exhibits transit depths of at most 3 -- 4\% in excess of their geometric transit. While at older ages, Ly$\alpha$ absorption is still significant (and possibly saturated for the lower mass planet), the H$\alpha$ absorption nearly disappears. This is because the extended atmosphere of neutral hydrogen becomes predominantly in the ground state after $\sim 1.2 $ Gyr. \end{abstract}
\begin{keywords}
hydrodynamics -- planets and satellites: atmospheres -- planetary systems -- methods: numerical
\end{keywords}

\section{Introduction}\label{sec.intro}
Following the first detection of an exoplanetary atmosphere around the hot-Jupiter HD209458 b  \citep{Charbonneau2001}, there has been a plethora of studies aimed at better understanding the atmospheres of exoplanets. Whether a planet can retain an atmosphere for a significant time period is essential for its habitability, as atmospheres regulate the planetary surface temperature, as well as absorb harmful solar or cosmic rays \citep[e.g.,][]{1993Icar..101..108K, 2009A&ARv..17..181L}. Consequently, studies regarding the stability of planetary atmospheres are critical in the ongoing searches for habitable planets. 

A planet that is too close to its host star, however, receives intense high-energy (i.e., X-rays and ultraviolet, UV) irradiation, which can cause atmospheres to heat, expand and `evaporate'. For many years now, the Hubble Space Telescope (HST) has been a key instrument for observing escaping planetary atmospheres in the UV \citep[e.g.][]{2003Natur.422..143V}. Major advancements in the field can be expected with new missions soon to be launched, such as the Colorado Ultraviolet Transit Experiment \citep{2017SPIE10397E..1AF}, and the James Webb Space Telescope. Predicting prime candidates for future observations of escaping atmospheres is therefore important both to guide these observations as well as for interpreting their findings. 

Strong atmospheric escape has been detected in several close-in exoplanets so far using  transmission spectroscopy in the UV \citep[e.g.,][]{2003Natur.422..143V,2010ApJ...720..872F, 2014ApJ...786..132K, 2015Natur.522..459E, 2016A&A...591A.121B}.  In particular, as the composition of gas giants consists predominantly of hydrogen, their atmospheres show strong absorption in the Ly$\alpha$ line. Unfortunately, the centre of this line is contaminated in our geocorona and can be significantly absorbed by neutral hydrogen in the interstellar medium \citep{2002ApJ...574..412W, Ben-Jaffel2008}. This renders the line core unusable and the signatures of escaping atmospheres can only be observed in the wings of the Ly$\alpha$ line.  

Another promising line for conducting transmission spectroscopy is H$\alpha$,  as this line can be observed with ground-based spectrographs and it does not suffer contamination at the line centre, thus allowing us to probe the accelerating material at the bottom of the escaping atmosphere. However, as the number of hydrogen atoms initially in the first excited state ($n=2$) can be much lower than the population at ground level ($n=1$), transit depths in H$\alpha$ are expected to be smaller than in Ly$\alpha$ transits \citep{2012ApJ...751...86J, 2015ApJ...810...13C}. Another complication is that an active host star shows variability in H$\alpha$, making it harder to disentangle signatures from the star and from the planetary atmosphere \citep{2017AJ....153..217C}. Recently, escaping atmospheres were detected using the metastable helium triplet at $\sim 10830$\AA\ \citep{2018Sci...362.1388N,2019A&A...623A..58A, 2019arXiv190713425A}, opening up new possibilities for probing planetary atmospheres with ground-based instrumentation in the infrared, such as with CARMENES \citep{2014SPIE.9147E..1FQ} and SpIRou  \citep{2018AJ....155...93C}. 

The reason why atmospheres of close-in giants are more susceptible to strong evaporation relies on two factors: 1) the intense high-energy irradiation they receive from their host stars, due their close proximity to the host, and 2) on their relatively low gravitational potential, required for a planet to `hold on' to their atmospheres. Young exoplanets are more susceptible to escape, because young stars have higher  X-ray and UV fluxes \citep{2005ApJ...622..680R, 2015ApJ...815L..12J} and because planets at young ages are more `puffed up', and thus have a lower gravitational potential. As the system evolves and the high-energy flux decreases with time, escape rates are expected to decrease. 

In this paper, we model the evolution of atmospheric escape on close-in giant planets. In particular, we focus on two different types of gas giants: a Jupiter-mass planet and a Saturn-mass planet (1 and 0.3 $M_{\rm jup}$, respectively). In both cases, we consider these planets to be at an orbital distance of 0.045 au or, equivalently, at a $3.5$~day orbit around a solar mass star. This distance is just inside the short-period `Neptunian desert', which shows a lack of planets with masses between 0.03 and 0.3 $M_{\rm jup}$ at orbital periods $\lesssim 5$--$10$ days  \citep{2016A&A...589A..75M}. The paucity of detected planets within this desert is suggestive of significant mass loss in planets with masses $\lesssim 0.3 ~M_{\rm jup}$. This will indeed be demonstrated in our results. 

Our paper is presented as follows. Section \ref{fEUVandrads} presents the two main inputs required in our study of the evolution of atmospheric escape: a prescription for how the high-energy flux of the host star evolves during the main-sequence phase and a prescription describing the contraction of the planet after its formation. These two factors are then considered in our atmospheric escape models, which are described in Section \ref{comp}, along with a description of the ray tracing technique used to simulate the spectroscopic transits. Our results are shown in Section \ref{evol_results_1}: for each age ranging from 10 to 5000~Myr, we  calculate the physical properties of the escaping atmospheres and predict their Ly$\alpha$ and H$\alpha$ transit profiles, which can be compared to observations. Section \ref{CONCLUSIONS} presents a discussion of our results and our conclusions.

\section{Evolution of planetary radii and stellar EUV Flux} \label{fEUVandrads}
\begin{figure}
 \centering
\includegraphics[width=0.4\textwidth]{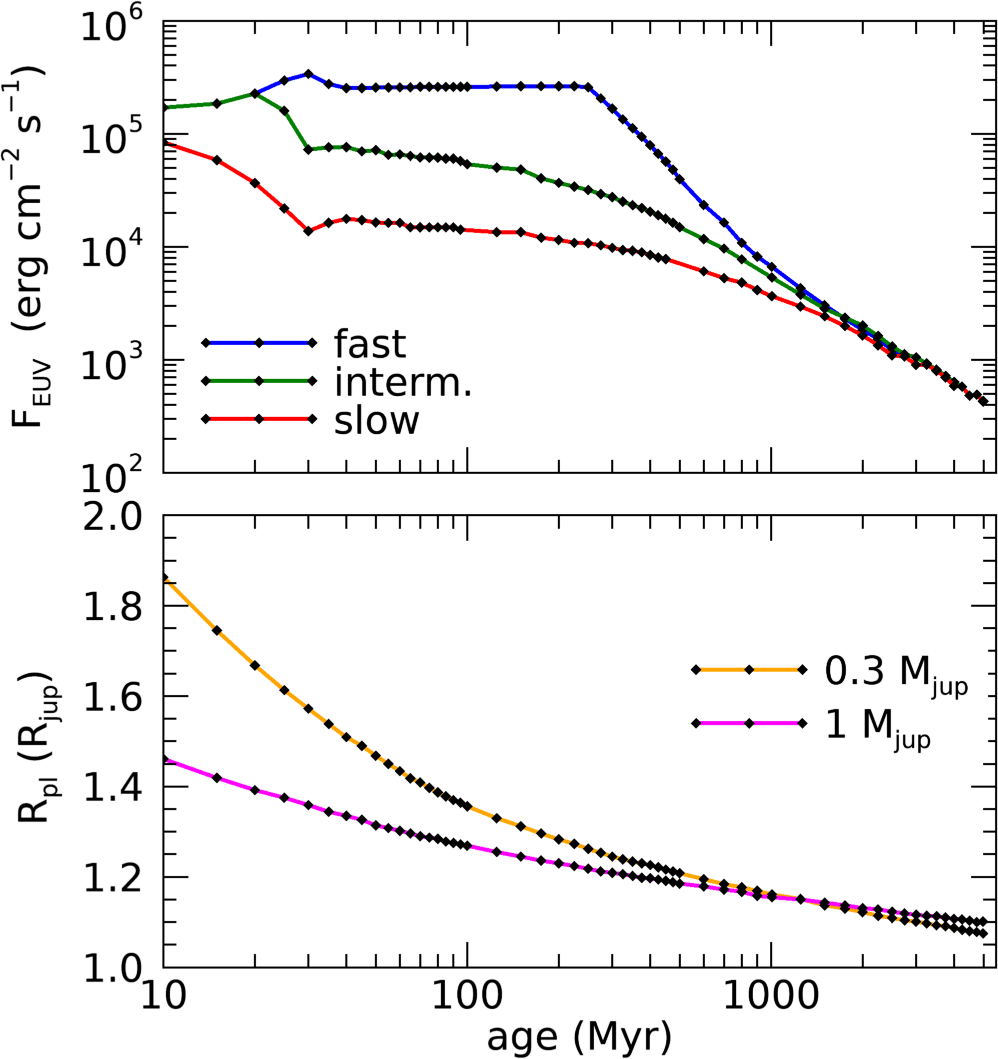}
\caption{Top: EUV flux received at an orbital distance of 0.045 au from a solar-like star with respect to planetary age. Line colour indicates stellar rotation: slow (red), intermediate (green), fast (blue). From \citet{Johnstone2015a}. Bottom: Planetary radius with respect to age for both a 1-$M_{\rm jup}$ (magenta) and a 0.3-$M_{\rm jup}$ (orange) close-in giant orbiting a solar-like star at 0.045 au. From \citet{Fortney2010}. The black diamonds show our sampled increments. }
\label{fig.fEUVandrad}
\end{figure}

The evolution of atmospheric escape of a close-in planet depends on two important factors: 
\begin{enumerate}
\item as the host star evolves, its activity declines due to spin down, resulting in declining fluxes in the extreme ultraviolet (EUV) and
\item as the planet evolves, cooling causes it to contract with time.
\end{enumerate}
 Therefore, to simulate the evolution of a close-in giant, we couple evolving predictions of EUV flux \citep{Johnstone2015a}  and planetary radius \citep{Fortney2010} with our hydrodynamic escape model, which will be described in Section \ref{comp}.  The evolution of these input parameters are shown in Figure \ref{fig.fEUVandrad} and described next. 
 
The EUV emitted from the host star originates from its magnetically heated, chromospheric and coronal plasma \citep[e.g.,][]{jardine}. Given that magnetic activity is linked to rotation, the EUV flux depends on the host star's rotational evolution. In rotational evolution studies, it is common to separate stars in different rotational tracks: those representing stars born as slow, intermediate or fast rotators \citep{2013A&A...556A..36G,2015A&A...577A..28J}. Due to the rotation-activity relations, stars that are born as slow rotators have overall lower EUV flux throughout their lives than those that were born as fast rotators \citep{2015A&A...577L...3T}. As the host star evolves, its magnetic activity declines due to spin down \citep{2014MNRAS.441.2361V}. At $\sim$ 1 Gyr, the rotational evolution paths for solar-mass stars converge to a unique solution \citep{1972ApJ...171..565S, 2013A&A...556A..36G}. This convergence is thus also predicted to occur in the evolution of the EUV flux ($F_{\rm EUV}$).  

Here, similarly to \citet{Johnstone2015a}, we consider three host star rotational evolutions, resulting in three differing EUV flux evolutions for fast, intermediate and slow rotators (Figure \ref{fig.fEUVandrad}, top). This allows us to study the influence of initial conditions of the star's rotation on atmospheric escape of an orbiting close-in giant.\footnote{The values predicted by \citet{Johnstone2015a} results in $F_{\rm EUV}=$ 6.7 erg s$^{-1} {\rm cm}^{-2}$ at 1 au assuming a solar-like star at the approximate solar age. To match the observed flux of 1 erg s$^{-1}{\rm cm}^{-2}$ from \citet{2005ApJ...622..680R}, we normalise the curves in \citet{Johnstone2015a} by dividing them by 6.7. The flux of 1 erg s$^{-1}{\rm cm}^{-2}$ is also the one adopted in the fiducial case of \citet{Murray-Clay2009}, on which our hydrodynamic model is based.}


 A planetary evolution model is required to calculate the change in planetary radius $R_{\rm pl}$ with time. Here, we use the results from \citet{Fortney2010} to describe the evolution of a 1-$M_{\rm jup}$ planet (magenta line in the bottom panel of Figure \ref{fig.fEUVandrad}) and of a 0.3-$M_{\rm jup}$ planet (orange line). These curves represent the calculations with no core in  \citet{Fortney2010}, for planets orbiting a solar-like star at 0.045 au, and they do not account for the mass loss. Ideally, as the atmosphere is removed from the planet, one would also need to take into account the reduction in the total mass of the planet, and consequently the corresponding change in radius. For example, this could be done by coupling atmospheric losses  to planetary evolution models  in a self-consistent way. This is left for a future work. Although mass loss is not self-consistent considered in the adopted radius evolution curves,  it is unlikely that this would affect the  1-$M_{\rm jup}$ planet, as total mass lost represents a small fraction of the total planetary mass, as we will see further on. However, this correction is more relevant for the  0.3-$M_{\rm jup}$, as lower-mass planets have stronger evaporation. We will discuss this further in Section  \ref{evol_results_1}. 

\section{Atmospheric escape and transit calculations}\label{comp}

\subsection{Hydrodynamic escape model} \label{WindmodelSection}
Because of the strong mass-loss rates inferred in observations ($> 10^9$g/s, \citealt{2011A&A...529A.136E}), atmospheric escape in close-in planets can be described using a fluid approximation. For escape to be considered  in this regime, the atmosphere has to be collisional, whose indicator is given by the Knudsen number, $K_n={\lambda_{\rm mfp}}/{h}$, which is the ratio between the mean free path of an atmospheric particle  $\lambda_{\rm mfp}$ and the atmospheric scale height $h$. The motion of a planetary atmosphere for which $ K_n \ll 1 $ is best described by a fluid, in an escape process known as hydrodynamic escape. This escape process occurs when heating in the collisional region of an atmosphere causes an upward pressure gradient force, which drives a bulk, radial outflow \citep{catling_kasting_2017}. 

Our numerical model is based on the model presented in \citet{Murray-Clay2009}. We treat the escaping atmosphere as a fluid, solving the equations of fluid dynamics in a co-rotating frame. These coupled differential equations are ionisation balance as well as the conservation of mass, energy and momentum. To achieve convergence of the wind solution, we use a shooting method based on the model of \citet{2006ApJ...639..416V}. 

In steady state, the momentum equation in spherical symmetry can be written as 
\begin{equation}\label{eq:con_mom}
{u}\frac{d {u}}{d r } = -\frac{1}{\rho}\frac{d P}{d r}   -    \frac{G M_{ \rm pl}}{r^2}+\frac{3GM_*r}{a^3}  ,
\end{equation}
where $r$ is the radial coordinate from the centre of the planet, $\rho$ and $u$ represent the atmospheric mass density and velocity, respectively, and $P$ is the thermal pressure. The first term on the right side of Equation (\ref{eq:con_mom}) is  the thermal pressure gradient while the second term represents attraction due to gravity. This equation is analogous to the momentum equation for a stellar wind \citep{1958ApJ...128..664P} with an additional term on the right side due to tidal effects. The tidal term is the sum of the centrifugal force and differential stellar gravity along the ray between the planet and star \citep{muno}. In this paper, we use the terms `hydrodynamic escape' and `planetary wind' interchangeably.

Conservation of energy requires that
\begin{equation}
    \rho {u} \frac{d}{d r} \left[\frac{k_BT}{(\gamma - 1)m} \right] = \frac{k_BT}{m}{u} \frac{d\rho}{d r }+Q-C  ,
\end{equation}
where $k_B$ is the Boltzmann constant, T the temperature, $\gamma = {5}/{3}$ is the ratio of specific heats for a monatomic ideal gas and $m$ is the mean particle mass, which ranges from $0.5m_H$ to $1m_H$, for a fully ionised to fully neutral atomic hydrogen plasma, respectively. Here, $m_H$ is the  mass of atomic hydrogen. The term on the left indicates the change in the internal energy of the fluid. The first term on the right represents cooling due to gas expansion, while the second ($Q$) and third ($C $) are heating and cooling terms. This equation is, again, typically used in stellar wind models, although with different heating and cooling terms \citep{2006ApJ...639..416V}.

Our planetary wind model assumes that the EUV flux is concentrated at one photon energy $e_{\rm in}=20$ eV, thus the volumetric heating rate [erg/s/cm$^3$] can be written as $Q=\epsilon F_{\rm EUV}e^{-\tau} \sigma_{\nu_0}n_{n}$, where $F_{\rm EUV}$ is the EUV flux received by the planet, $\tau$  the optical depth to ionising photons, $n_n$  the number density of neutral hydrogen and $\sigma_{\nu_0}$ is the cross section for the photoionisation of hydrogen  given by 
$\sigma_{\nu_0}=6\times 10^{-18} \left( {e_{\rm in}}/{13.6 {\rm eV}}\right)^{-3}=1.89 \times 10^{-18}{\rm cm}^2$
\citep{spit}. Of the total received energy flux $F_{\rm EUV}$, a fraction $\epsilon$ is converted to thermal energy to heat the atmosphere. In our simulations, we follow the approach by \citet{Murray-Clay2009} and assume $\epsilon= 1- {13.6 ~\rm{eV}}/{e_{\rm in}}=0.32$.

We assume that the escaping atmospheric gas cools by radiative losses resulting from collisional excitation, also known as Ly$\alpha$ cooling. The volumetric cooling rate  is $ C= 7.5 \times 10^{-19} n_+ n_n \exp[{-1.183\times 10^5}/{T}]$, where $n_+$ is the number density of protons, which is equivalent to the number density of electrons in a purely hydrogen plasma. 

The ionisation balance  is given by
 \begin{equation}\label{eq: ion_bal}
\frac{n_n F_{\rm EUV}e^{-\tau}\sigma_{\nu_0}}{e_{\rm in}}= n_+^2\alpha_{\rm rec}+\frac{1}{r^2}\frac{d}{d r}(r^2n_+{u})  , 
\end{equation}
where  the rate of photoionisations (first term) balances the sum of the rates of radiative recombinations (second term) and of the advection of ions (third term). Here, $\alpha_{\rm rec} = 2.7 \times 10^{-13} (T/10^4)^{0.9}$  is the case B radiative recombination coefficient for hydrogen ions \citep{95, 2006agna.book.....O}. The ionised fraction is given as $F_{\rm ion}={n_+}/{n_H}$, where $n_H=n_+ + n_n$ is the total number density of hydrogen.

Finally, the conservation of mass requires that 
\begin{equation}\label{eq: cons_mass_1st}
\frac{d (r^2 \rho {u}) }{d r } = 0  . 
\end{equation}
Our simulated planetary outflow originates from the sub-stellar point of the planet, which is the point closest to the star. We then apply our calculated solution over 4$\pi$ steradians rendering it an upper limit to atmospheric escape \citep{Murray-Clay2009, 2015ApJ...815L..12J}. Therefore, from Equation (\ref{eq: cons_mass_1st}), we have that the mass-loss rate of the escaping atmosphere is $\dot{M} = 4\pi r^2 \rho {u} $.

Similarly to stellar wind theory, the initially subsonic flow is accelerated transonically until it reaches an asymptotic speed, known as the terminal velocity $u_{\rm term}$. Convergence is reached once the values of the mass-loss rate and terminal velocity between two subsequent runs are below 1\%. We then calculate the Knudsen number $K_n$ for each run of our simulations, so as to ensure that the atmosphere remains collisional throughout all our simulations.

\begin{figure}
\centering
\includegraphics[width=0.47\textwidth]{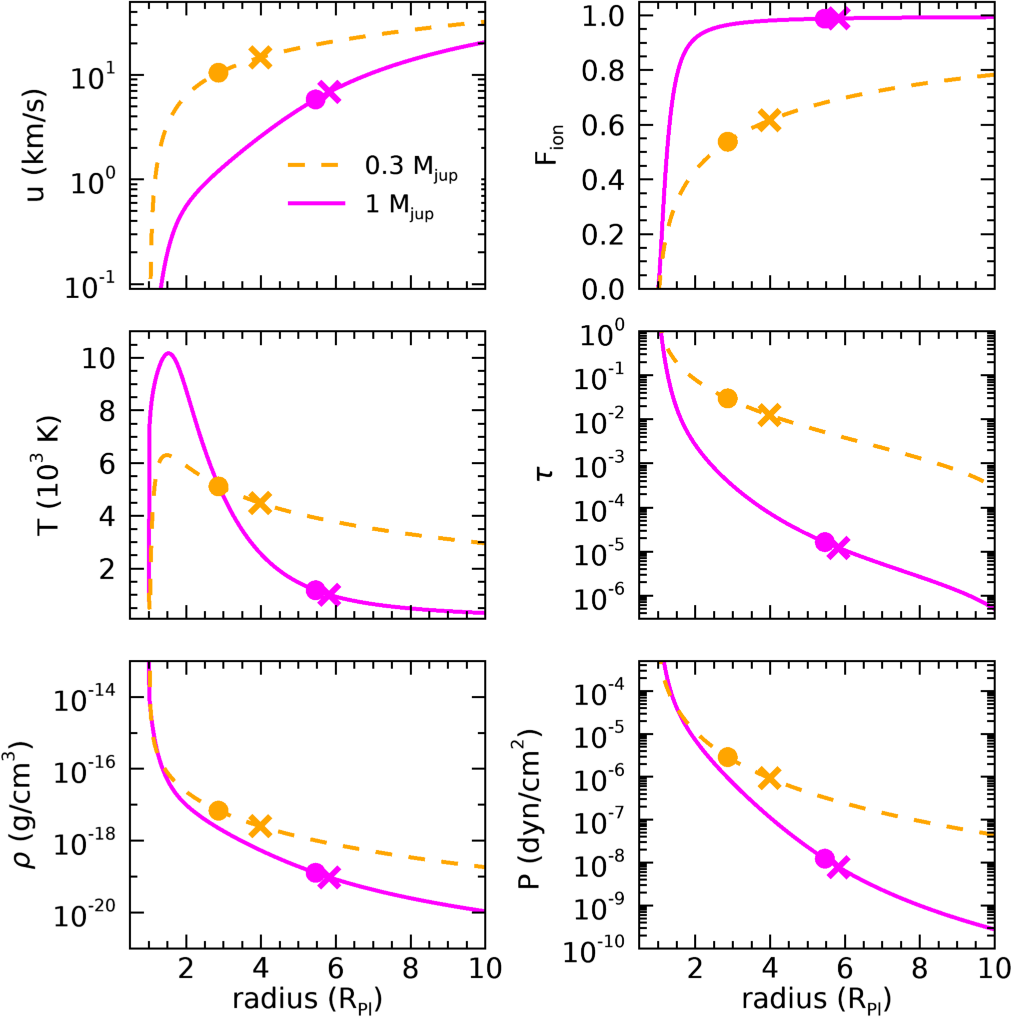}
\caption{Radial profiles of various atmospheric properties for two simulated close-in planets with masses $1~M_{\rm jup}$ (magenta, solid line) and $0.3M_{\rm jup}$ (orange, dashed line) and radius $1.1~R_{\rm jup}$. The planets orbit a solar-like star at 0.045 au, and receive an incident flux $F_{EUV} = 480$ erg cm$^{-2}$ s$^{-1}$, consistent with a 4.5 Gyr-old star.  Left panels, from top to bottom: wind velocity $u$, temperature $T$,  total mass density $\rho$. Right panels, from top to bottom: fraction of ionised hydrogen $F_{\rm ion}$, optical depth to ionising photons $\tau$ and thermal pressure $P$. The distance at which $\tau=1$ is reached at $\sim 1.1 R_{\rm pl}$ for the cases presented above. The circle and cross indicate the sonic point and Roche lobe boundary, respectively. 
\label{fig:fiducial}}
\end{figure} 

Figure \ref{fig:fiducial} shows typical solutions of our simulations, where we show radial profiles of various atmospheric parameters for  $1~M_{\rm jup}$ (magenta) and  $0.3~M_{\rm jup}$ (orange) planets orbiting a solar mass star at 0.045 au. The adopted radii for these illustrative simulations are $1.1~ R_{\rm jup}$. For both these planets, we assume an $F_{\rm EUV} = 480$ erg cm$^{-2}$ s$^{-1}$. In addition to these parameters, which determine the physical properties of the planetary systems, our hydrodynamic models require values of the temperature  $T_0$ and density $\rho_0$ at the base of the escaping atmosphere (here, assumed at 1 $R_{\rm pl}$). For all of our simulations, we assume these parameters are $T_0=1000$ K and  $\rho_0=4 \times 10^{-13}$ g cm$^{-3}$, similar to values adopted in \citet{Murray-Clay2009}. Note that, although  $T_0$ and  $\rho_0$ are free parameters of the model, \citet{Murray-Clay2009} demonstrated that large variations in these values had negligible effect on the resulting simulated escape. We confirmed their results with our models as well.  

The left panels of Figure \ref{fig:fiducial} show profiles of planetary wind velocity $u$, temperature $T$, and  total mass density $\rho$, while the right panels  show  fraction of ionised hydrogen $F_{\rm ion}$, optical depth to ionising photons $\tau$ and thermal pressure $P$. In each of the panels, the cross represents the Roche lobe boundary 
\begin{equation}
R_{\rm Roche} = \frac{a}{R_{\rm pl}}\left(\frac{M_{\rm pl}}{3M_{*}}\right)^{1/3} .
\end{equation}
Atmospheric particles are no longer gravitationally bound to the planet for $r>R_{\rm Roche}$, with stellar tidal gravity dominating over planetary gravity. Note that each of our calculations extend out to 10 $R_{\rm pl}$, which is above $R_{\rm Roche}$.  The circle indicates the sonic point $R_{\rm sonic}$, which is the distance at which the planetary wind reaches the sound speed $c_s = \sqrt{{5 k_BT}/{ 3 m }}$. It is evident from the velocity profile that our simulated wind becomes transonic within the Roche lobe boundary. This is also the case for our simulated planets featured in the subsequent sections during all evolutionary stages. 

Figure \ref{fig:fiducial} shows a steep drop in density and pressure with simultaneous sharp rises in $T$ and $F_{\rm ion}$, occurring below $1.5 R_{\rm pl}$. These strong variations in the atmospheric structure are the result of strong absorption of stellar EUV radiation at this distance (region with high optical depth $\tau$), which leads to the photoionisation of atmospheric hydrogen, as seen in the dramatic increase of $F_{\rm ion}$ from $\simeq 0$\% to $75\%$ ($30\%$) at $1.5 R_{\rm pl}$ for a 1-$M_{\rm jup}$ (0.3-$M_{\rm jup}$) planet. Ejected free electrons then heat the planetary atmosphere through collisions, causing the sharp temperature increase. The high temperatures cause the atmosphere to expand and escape hydrodynamically.  
At large distances, the temperature drops due to the expansion of the atmosphere (adiabatic cooling). This energy sink dominates over heating once the atmosphere becomes sufficiently ionised. Ly$\alpha$ cooling also acts against the heating and is at its strongest at higher atmospheric temperatures due to increased collisional excitation \citep{Salz2016b}. 

Comparing the curves for the 1-$M_{\rm jup}$ and 0.3-$M_{\rm jup}$ planets in Figure \ref{fig:fiducial}, we see that the wind of the more massive planet has lower velocities due to its higher gravitational potential. Its atmosphere is also more rarefied, as seen by the density and optical depth curves. Its temperature, nevertheless, reaches higher values, leading to a more ionised atmosphere. The mass-loss rates of the 1-$M_{\rm jup}$ planet is $1.7  \times 10^{9}$ g s$^{-1}$, 26 times lower than the mass-loss rate of $4.4 \times 10^{10}$ g s$^{-1}$, for the 0.3-$M_{\rm jup}$ planet. Altogether, the lower gravitational potential of the less massive planet results in stronger planetary winds, with higher velocities and mass-loss rates.

\subsection{Comparison to energy-limited escape}
The energy limited approximation  \citep{Wats} is often adopted for calculating mass-loss rates of atmospheric escape in place of more computationally heavy hydrodynamic  models. 
This approximation can be derived as follows. We first assume energy balance between the final kinetic energy of the planetary wind and the input energy due to irradiation
\begin{equation} 
E_{\rm kinetic, output} = E_{\rm irradiation, input}
\end{equation}
For the input energy,  a fraction $\epsilon$ of the stellar high-energy flux $F_{\rm EUV} $ intercepted by  a planet with a cross section of $\pi R_{\rm pl}^2$ is assumed to drive a flow that reaches a kinetic energy $ \dot{M} u_{\rm term}^2/2$, where $u_{\rm term}$ is the terminal velocity.
\begin{equation} \label{eq.el}
\frac{\dot{M} u_{\rm term}^2}{2} = \epsilon F_{\rm EUV} (\pi R_{\rm pl}^2) \, .
\end{equation}
where $\dot{M}$ is the mass-loss rate of the planetary wind. We further assume that the terminal velocity $u_{\rm term}$ is on the order of the surface escape  velocity $v_{\rm esc} = (2 G M_{\rm pl}/R _{\rm pl})^{1/2}$, we have thus that the mass-loss rate derived in the energy-limit approximation is
\begin{equation} \label{eq: E_lim_simple}
\dot{M}_E = {\epsilon} \frac{ F_{\rm EUV} (\pi R_{\rm pl}^2)  }{ GM_{\rm pl}/R _{\rm pl}} .
\end{equation}

Using an efficiency of $\epsilon=0.32$, for the 1-$M_{\rm jup}$ and 0.3-$M_{\rm jup}$ planets presented in Figure 1, we find energy limited mass-loss rates of   $\dot{M}_E = 1.9 \times 10^9$ g s$^{-1}$ and $6.2 \times 10^{9}$ g s$^{-1}$, respectively. 
The energy-limit approximation overestimates the mass-loss rate computed from hydrodynamical models for the  1-$M_{\rm jup}$ planet by only a factor of 1.1. However, it underestimates by a factor of 1/7 the mass-loss rate of the 0.3-$M_{\rm jup}$ planet. Examining Equations (\ref{eq.el}) and (\ref{eq: E_lim_simple}), we expect that $\dot{M}_E > \dot{M}$, when the terminal velocities of the wind are  $u_{\rm term} < v_{\rm esc}$. Indeed, for the  1-$M_{\rm jup}$ planet, terminal velocities are $\sim 20$~km/s, compared to an escape velocity of $\sim 60$~km/s. On the other hand, we expect that $\dot{M}_E < \dot{M}$, when the terminal velocities of the wind are  $u_{\rm term} > v_{\rm esc}$. This is the case for the 0.3-$M_{\rm jup}$ planet, where the terminal velocities are above 32~km/s and escape velocity is $\sim 30$~km/s.

Recent studies showed that the energy-limited approximation can be a poor indicator of the mass lost in atmospheric escape processes \citep[e.g.][]{Salz2016,kub2} and can underestimate the mass-loss rate of low density, highly irradiated exoplanets planets in the boil-off regime  \citep{kub2}. To better reconcile the analytical expression with hydrodynamic models, a correction factor $K$ is often considered in the denominator of Equation (\ref{eq: E_lim_simple})  to account for  stellar tidal forces that reduces the effect of the planetary gravitational potential \citep{Erkaev2007}
\begin{equation}
K=1-\frac{3R_{\rm pl}}{2R_{ \rm Roche}}+\frac{R_{\rm pl}^3}{2R_{ \rm Roche}^3}  .
\end{equation}
Additionally, the cross section of the planet is assumed to be $\pi R_{\tau=1}^2$, where $R_{\tau=1}$ refers to the distance where optical depth unity to ionising photons is reached. These two additional factors can reduce the discrepancy between the hydrodynamic calculations and the analytical limit, hence
\begin{equation} \label{eq: E_lim}
\dot{M}_{E}^{\rm corr} = {\epsilon} \frac{ F_{\rm EUV} (\pi R_{\tau=1}^2)  }{K GM_{\rm pl}/R _{\rm pl}} .
\end{equation}
However, to calculate $R_{\tau=1}$, one needs to compute the density and optical depth profiles through hydrodynamic simulations or from a derived prescription based in such simulations. For our examples presented in Figure \ref{fig:fiducial}, using $R_{\tau=1}$ from our simulations, and the tidal correction factor $K$, we find $\dot{M}_{E}^{\rm corr}$ that are factors of 1.6 and 0.25 the values from our hydrodynamical models. In these cases, the `correction' made the discrepancy smaller for the 0.3-$M_{\rm jup}$ planet (from 1/7 to 1/4), but increased the discrepancy for the 1-$M_{\rm jup}$, when we adopted Equation (\ref{eq: E_lim}) instead of (\ref{eq: E_lim_simple}). In the comparisons done further on in this paper, we compute energy-limited mass loss from Equation (\ref{eq: E_lim}).

\subsection{Transit calculations} \label{Raymodel}
After solving for the hydrodynamic equations for the planetary outflow, we use a ray tracing model to simulate planetary transits as observed in  both Ly$\alpha$ and H$\alpha$. This model is described in \citet{2018MNRAS.481.5296V}, here modified for Ly$\alpha$ and H$\alpha$ transitions. In the ray tracing model presented here, we  upgrade from a Doppler line profile to a Voigt line profile. As the ray tracing model from \citet{2018MNRAS.481.5296V} is three dimensional, we create a three-dimensional Cartesian grid with 201 evenly spaced  cells in each direction. This grid is centred on the planet and each cell is `filled'\footnote{Similar to a solid of revolution.} with the one dimensional hydrodynamic calculations of $u$, $T$, and $\rho$, taking into account that Doppler shifts result from the projection of the wind velocities along the line-of-sight. One of the axis of the grid is aligned along the observer-star line, so that the grid seen in the plane-of-the-sky is a square of $201 \times 201$ cells. 

To calculate a frequency-dependent transit, we use 51 velocity channels from $u_{\rm channel}=-500$~km/s to $+500$ km/s, using increments of 50 km/s for the outer velocities ($|u_{\rm channel}|=100$ to $500$ km/s), while we improve the resolution at  line centre (35 channel increments within $\pm$ 100 km/s). 
With this, the frequency-dependent optical depth along a single ray in the direction connecting the observer to the star-planet system is
 \begin{equation}
 \tau_\nu  =  \int n_{n} \sigma \phi_\nu dz  , 
 \end{equation}
where $\sigma$ is the absorption cross-section at the line centre,  $\phi_\nu$  the Voigt line profile. The subscript `$\nu$' indicates that the line profile and the optical depth are dependent on the frequency  (or velocity channel) of the observation. We calculate $\sigma$ for both Ly$\alpha$ and H$\alpha$ transitions using 
\begin{equation}
\sigma =  \frac{\pi e^2 f}{m_e c }
\end{equation}
 where $f$ is the oscillator strength of the transition, $m_e$ the electron mass, $e$ the electron charge and $c$ the speed of light. Our values of $f$ for each of these transitions were obtained from the NIST Catalogue\footnote{ \url{https://www.nist.gov/pml/atomic-spectra-database}}, which gives $f=0.416410$ for Ly$\alpha$ and $f=0.64108$ for H$\alpha$. The Voigt line profile $\phi_\nu$ is a convolution of a Gaussian and Lorentzian line profile accounting for both Doppler and natural broadening
\begin{equation}
\phi_\nu =\frac{\lambda_0}{\sqrt{\pi}u_{\rm th}}  \frac{\xi}{\pi}\int_{-{\infty}}^{{\infty}}\frac{e^{-w^2}}{\chi^2 +(\Delta {u}/u_{\rm th}-w)^2 }dw
\end{equation}
where $\lambda_0 $ is the wavelength at line centre ($\lambda_0 = 1215.67 $\AA\ for Ly$\alpha$; $ \lambda_0 = 6562.79 $\AA\ for H$\alpha$),  $u_{\rm th} = (2 k_B T / m_H)^{1/2}$ and $m_H$ is the  mass of atomic hydrogen. $\chi={A_{ji} \lambda_0}/({4 \pi u_{\rm th}})$ is the damping parameter, where $A_{ji}$ is the transition rate ($A_{ji}=4.69860 \times 10^8$ s$^{-1}$ for the Ly$\alpha$ transition; $A_{ji}= 4.4101 \times 10^7$ s$^{-1}$ for the H$\alpha$ one; values from NIST).  The velocity offset from the line centre is $\Delta {u} = {u_{\rm channel}}-{u}_{\rm LOS}$,  where ${u}_{\rm LOS}$ is the line of sight flow velocity of the escaping wind. We make use of IDL's inbuilt \texttt{voigt} function to calculate  $\phi_\nu$. 

In computing the transmitted spectrum, we neglect  centre-to-limb variations in the stellar disc and assume that, for a given frequency, the stellar disc emits a uniform specific intensity $I_\star$. The fraction of `transmitted' specific intensity at a given frequency  during transit is hence 
\begin{equation}
\frac{I_\nu}{I_\star} = e^{-\tau_\nu} .
\end{equation}
$1 - I_\nu/{I_\star} $, therefore, represents the fraction of absorbed specific intensity as a result of the absorption from both the planet disc and its atmosphere. To calculate the total absorption, we then integrated over all rays. Given our three-dimensional grid construction, the grid projected in the plane of the sky has $201 \times 201$ cells. We therefore shoot 51 frequency-dependent rays through each of the $201^2$ grid elements. Given that the grid is larger than the projected area of the stellar disc, rays emitted from  pixels outside the stellar disc are assigned a zero specific intensity. 

Therefore,  integrating over all these rays and dividing by the `flux' of star ($\pi R_\star^2$) allows us to calculate the transit depth 
\begin{equation}\label{eq.deltaF}
\Delta F_\nu = \frac{\int\int (1-e^{-\tau_\nu})dxdy}{\pi R_\star ^2} .
\end{equation}
where $dx$ and $dy$ are the sizes of each cell. By construction, $dx=dy=2 \times 10R_{\rm pl}/201 \simeq 0.1R_{\rm pl}$ (the factor of 2 comes from the `solid of revolution' method used and $10R_{\rm pl}$ is the extension of the hydrodynamical computations). In practice, $\Delta F_\nu$  represents the ratio of a frequency-dependent effective area of the planet $\pi(R^{\rm eff}_{{\rm pl}, \nu})^2$, i.e., planet disc plus its absorbing atmosphere, by the stellar area
\begin{equation}
\Delta F_\nu =  \frac{\pi (R^{\rm eff}_{{\rm pl}, \nu})^2}{\pi R_*^2}. 
\end{equation}
Because of the nature of our spherically symmetric model, the effective area of the planet is that of a circle, but in 3D models, this area could take a different form \citep[e.g.][]{2018MNRAS.481.5296V}. Here, we assume a transit along the centre of the stellar disc resulting in an impact parameter $b=0$. Assuming circular orbit, this gives a transit duration of
\begin{equation}\label{eq: tran_dur}
t_{\rm dur}    \approx   \frac{P_{\rm orb}R_\star }{\pi a}  = \frac{2 R_\star}{\sqrt{G M_*/a}} = 2.75 ~ \rm h ,
\end{equation}
where $P_{\rm orb}$ is the orbital period.

\section{Evolution of Atmospheric Escape in close-in giants} \label{evol_results_1}

\begin{figure}
\begin{center} 
\hspace*{-0.5cm} 
\includegraphics[width=0.47\textwidth]{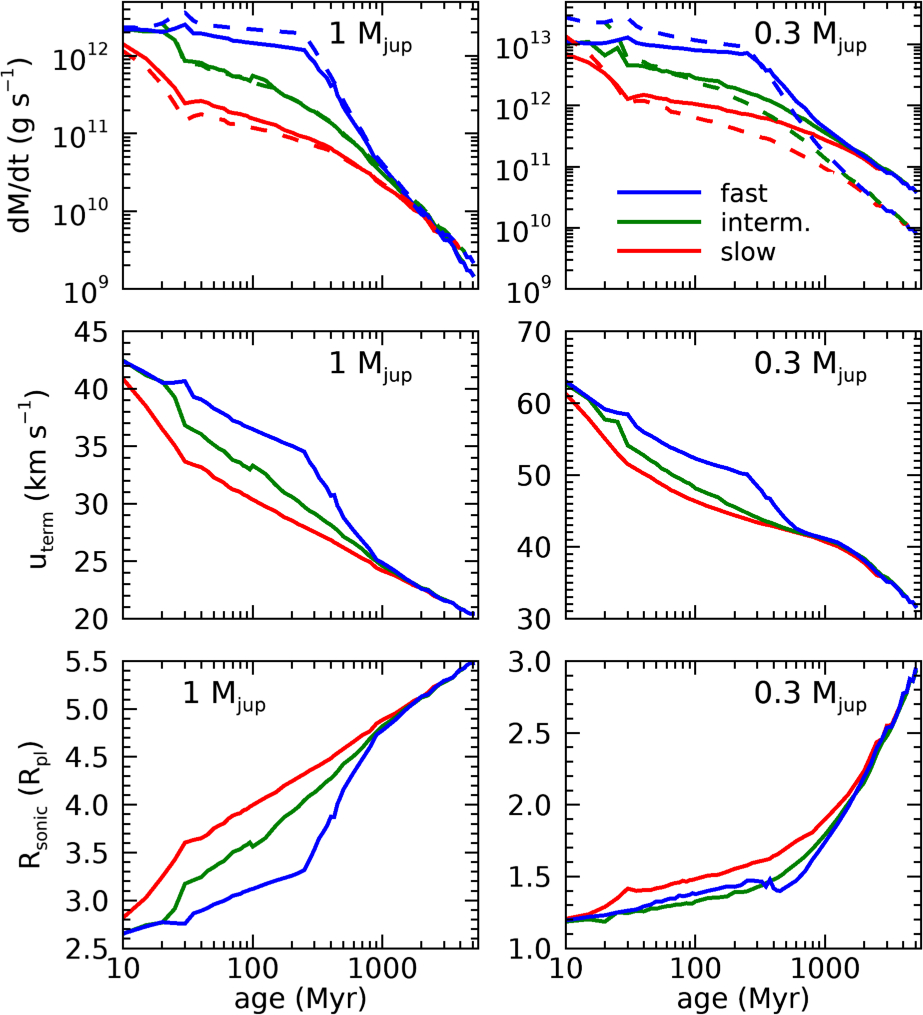}
\caption{Evolution of atmospheric properties for a 1-$M_{\rm jup}$ (left) and 0.3-$M_{\rm jup}$ (right) planets orbiting a 1-$M_\odot$ star at 0.045 au, of slow (red), intermediate (green) or fast (blue) rotation. The panels are: mass-loss rate $\dot{M}$ (upper), terminal velocity u$_{\rm term}$ (centre) and radial distance to the sonic point $R_{\rm sonic}$ (lower).  Dashed lines show $\dot{M}$ predicted by the energy-limited approximation (Equation \ref{eq: E_lim}). We present analytical fits to these results in Appendix \ref{evol_analt}.}
\label{fig.evol}
\end{center}
\end{figure}

To probe the evolution of the atmospheric escape, we perform escape simulations for two planet masses (0.3 and 1 $M_{\rm jup}$) orbiting stars that were born as slow, intermediate and fast rotators. This results in six different evolutionary tracks. For each track, we have more than 50 models sampling ages from 10~Myr to 5~Gyr. The ages sampled are shown as diamonds in Figure \ref{fig.fEUVandrad}. In total, we ran more than 300 models to compute the evolution of atmospheric escape. For each of these models, we also predict  Ly$\alpha$ and H$\alpha$ transit signals.
 The stellar mass ($M_*= 1 M_\odot$), radius ($R_* = 1 R_\odot$) and orbital distance ($a=0.045$~au) are fixed in our study.
In Figure \ref{fig.evol}, we present the evolution of three  atmospheric  properties predicted using our models: the mass-loss rate (top panels), terminal velocities (middle) and the position of the sonic point (bottom). The results for our 1-$M_{\rm jup}$ and 0.3-$M_{\rm jup}$ planets are shown  on the left and right panels, respectively. 

The evolution of $\dot{M}$ for the three considered host star rotations are shown as solid lines in the top panels of Figure \ref{fig.evol}. We see that close-in giants experience greater levels of mass loss during younger evolutionary stages and that their mass-loss rates decrease with time. This is the result of the lowered EUV flux incident on the planetary atmosphere as well as the contracting planetary radius with evolution (Figure \ref{fig.fEUVandrad}). The lowered EUV flux leads to a reduced level of heating in the atmosphere due to a lower number of photoionisations. This ultimately weakens the hydrodynamic wind as the planet evolves, resulting in the declining curves of $\dot{M}$.

The contracting planetary radius with evolution results in a strengthening planetary gravitational force on atmospheric particles. This acts to further reduce the atmospheric escape experienced with evolution by improving the planet's ability to retain its atmospheric particles. A weaker gravitational potential is directly responsible for the stronger escape experienced by the lower mass 0.3-$M_{\rm jup}$ planet, as compared to the  1-$M_{\rm jup}$ planet. Both its lower mass and larger radii are responsible for this. 

Although not considered in our models, the decreasing planetary mass resulting from atmospheric escape will soften the effect of the growing planetary gravity with age.  We do not include this decrease in mass in our simulations. As a result, our mass-loss rates increasingly underestimate the true mass-loss rate as the planet ages due to a growing overestimation of planetary gravity. Fortunately, we can assume that the resulting inaccuracy is of minor importance as the effect of the neglected mass variation on the planetary gravity is negligible in comparison to that resulting from the radius variation during evolution (Section \ref{evol_tot_mass}).

We used the evolutionary curves from  \citet{Fortney2010}, which did not consider the presence of a core. By considering a 25-$M_\oplus$ core, \citet{Fortney2010} predicted planets with smaller radii than those without a core. In our models, this smaller radius would lead to stronger planetary gravity and consequently to a somewhat weaker mass loss. However, this would not change the trends we present in Figure \ref{fig.evol}.

The significantly steeper $\dot{M}$ curves for the close-in giant orbiting a fast-rotating star (blue) further demonstrates the significance of stellar activity on atmospheric escape. Stellar rotation determines the EUV emission, with faster rotation producing higher emission, until a saturation plateau is reached for a fast rotating star (Figure \ref{fig.fEUVandrad}). The $\dot{M}$ curves for the three rotations become indistinguishable after $\sim 1$ Gyr, owing to a  convergence of the received EUV flux at this age.

The dashed lines in the upper panels of  Figure \ref{fig.evol} show the mass-loss rates estimated using the energy limit  approximation (Equation \ref{eq: E_lim}). For the 1-$M_{\rm jup}$ planet, the energy limited approximation overestimates the mass loss when orbiting a fast rotating host star (i.e., in the case of high $F_{\rm EUV}$), while it underestimates mass-loss rates in the case of planets orbiting slowly rotating hosts. The results for the hydrodynamic case and the energy-limit case agree quite well after $\sim 500$~Myr. For the case of the 0.3-$M_{\rm jup}$ planet, the situation is quite different, in which we see that the estimates of $\dot{M}$ provided by the energy-limited approximation are very different from the hydrodynamical calculation. 

The middle panels of Figure \ref{fig.evol} show the terminal velocity ${u}_{\rm term}$ of the planetary outflow, where we see that at younger ages, larger terminal velocities are reached. We define ${u}_{\rm term}$ here as the velocity of atmospheric particles leaving the planetary atmosphere at our maximum simulated distance of 10 $R_{\rm pl}$. As with the curves of $\dot{M}$, we see a decline of the ${u}_{\rm term}$ as the planet ages. However, while mass-loss rates vary by more than 3 orders of magnitude from 10 to 5000 Myr, the terminal velocities only vary by a factor of $\sim 2$.
Both $\dot{M}$ and ${u}_{\rm term}$ decline with age due to a combination of an increasingly weaker atmospheric heating and an improved ability of the planet to retain its atmospheric particles causing weaker, slower hydrodynamic winds at later evolutionary stages. The velocity of escaping neutral hydrogen atoms is of particular importance for Ly$\alpha$ observations as only absorption in line wings are observable (Section \ref{lya_sec}).

The bottom panels of Figure \ref{fig.evol} show the evolution of the sonic point  $R_{\rm sonic}$, which climbs to higher positions of the atmosphere with evolution. The sonic point occurs when the wind velocity reaches the speed of sound, $u = c_s$. With evolution, there is a slower rise in the wind velocity with distance (i.e., lower acceleration). As a result, the wind must travel a greater distance before it reaches the sound speed $c_s$. With age, $c_s$ will also be slightly lowered  due to overall lower atmospheric temperatures, but this only partially counteracts the effects of the lower wind acceleration. As a result, the variations in wind  velocity profiles dominate in setting the growth of $R_{\rm sonic}$ with evolution.

We draw attention to the striking resemblance of each of the  curves of Figure \ref{fig.evol} to the curves of the received $F_{\rm EUV}$ included in Figure \ref{fig.fEUVandrad}, demonstrating a strong dependence of $\dot{M}$ and ${u}_{\rm term}$ on the irradiation level of the planetary atmosphere. We present analytical fits to these results in Appendix \ref{evol_analt}.

\subsection{Total mass lost during evolution} \label{evol_tot_mass}
\begin{figure}
\begin{center} 
\includegraphics[scale=0.49]{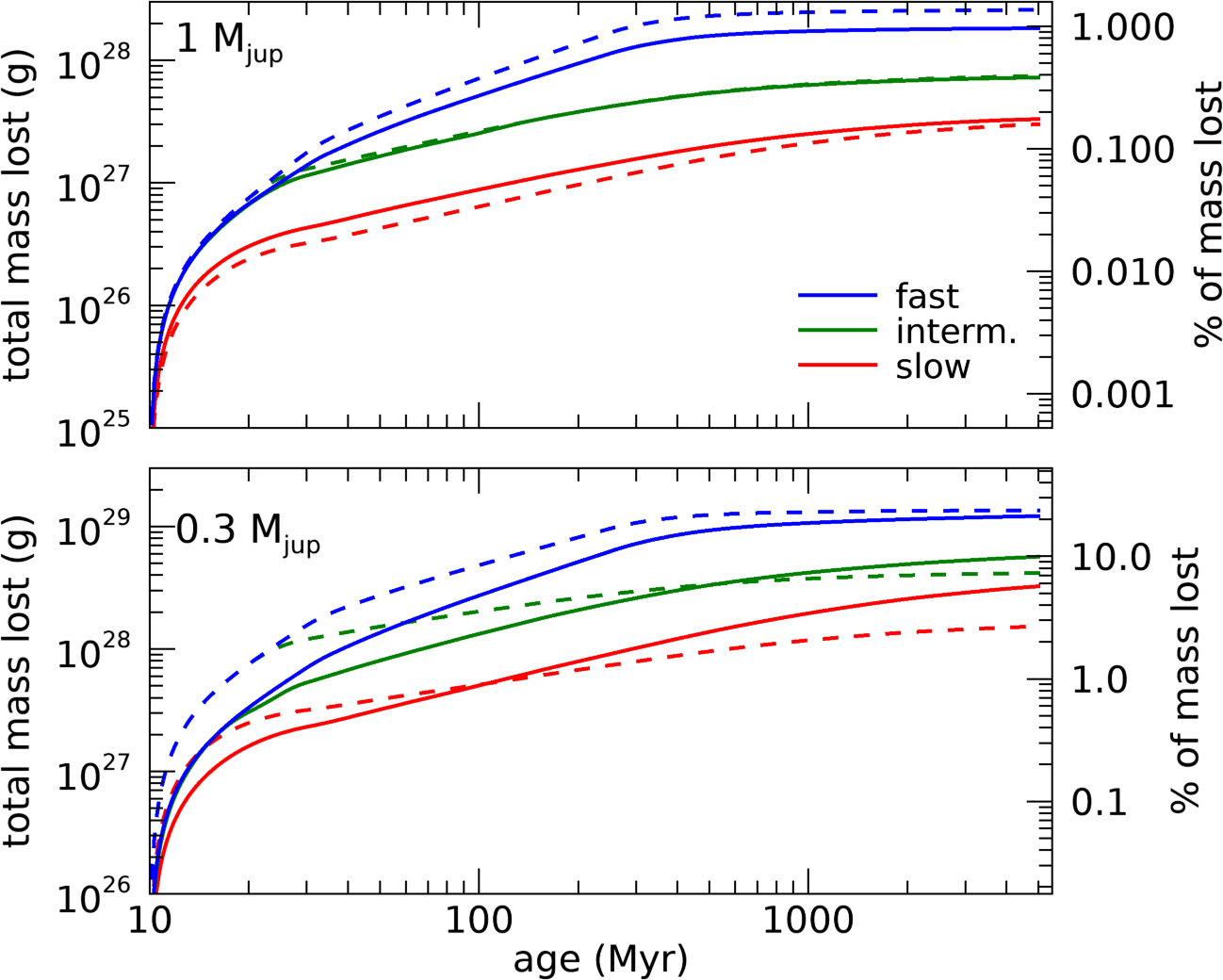}
\caption{Total planetary mass lost after the age of 10 Myrs for a 1- (top) and 0.3- (bottom) $M_{\rm jup}$ planets orbiting a slow (red), intermediate (green), and fast (blue) rotating host. Solid and dashed lines are mass lost calculated using the hydrodynamic  model and energy-limited case, respectively. }
\label{fig: totcurve}
\end{center}
\end{figure}

Figure \ref{fig: totcurve} shows the total mass lost calculated over the simulated life of the  1- (top panel) and 0.3-$M_{\rm jup}$ (bottom) planets. We calculate the total mass lost by integrating  mass-loss rates given in Figure \ref{fig.evol} with respect to time. Figure \ref{fig: totcurve} shows the largely differing mass losses resulting from the three considered rotations of the host star. The total mass lost is clearly dependant on stellar rotation and mass loss is greatest for planets orbiting faster rotating hosts. The mass loss experienced by each planet levels off with age due to the slowing mass-loss rates with evolution shown in Figure \ref{fig.evol}. 

Past studies predicted that atmospheric escape does not play an influential role in the overall evolution of hot Jupiters \citep{Murray-Clay2009,2013ApJ...775..105O,2015Natur.522..459E}. In spite of the large mass-loss rates observed, the planetary mass is still very large, constituting essentially an unending  atmospheric reservoir. 
Indeed, over its entire evolution, we predict that the 1-$M_{\rm jup}$ planet would lose at most $1\%$ of its  initial mass. We therefore expect hydrodynamic escape to have a negligible effect on mass of highly irradiated Jupiter-mass planets, owing to their large mass reservoir. 


However, for the 0.3-$M_{\rm jup}$ planet, our results indicate the opposite -- mass lost due to hydrodynamic escape is considerably large, and could potentially shape the planet's evolution. In this case, the total mass lost can reach more than 20\%. For orbital distances even shorter than the 0.045~au considered in this work, mass-loss rates are expected to increase even  more, giving support to the idea that close-in  rocky planets might have begun their evolution as Neptune-like planets, but eventually evaporated the entirety of their atmospheres \citep{2015Natur.522..459E}. 

Regardless of the planetary mass, orbiting a faster rotating host star leads to greater losses, with a slow rotator host producing  $\sim$ 4 to 5 times less lost mass than a fast rotating host. The dashed lines in Figure \ref{fig: totcurve} show the total mass lost calculated using the energy-limit approach. This approximation overestimates the total mass lost when orbiting a fast rotating host star, while it underestimates the loss for planet orbiting a slow host. 


\section{Spectroscopic transits in Hydrogen lines}\label{lya_and_ha}
We use the solutions of the hydrodynamic calculations presented in the previous Section to calculate spectroscopic transits in Ly$\alpha$ and H$\alpha$. To model these transits, we need to know the density of neutral hydrogen in the ground state $n=1$ (hereby $N_1$) and in the first excited state $n=2$ ($N_2$), respectively. 
For that, we solve the statistical equilibrium equation in the coronal-model approximation \citep{2018LRSP...15....5D}. In this approximation, the spontaneous radiative decay  balances the excitation process, and collisional de-excitation is neglected. Only direct excitations from the ground state  $n=1$ are included. To perform this calculation, we use the CHIANTI software \citep[v 9.0.1, ][]{2019ApJS..241...22D}\footnote{We use \texttt{sun\_photospheric\_2015\_scott.abund}  for the photospheric abundance file and ion equilibrium file \texttt{chianti.ioneq}.} written in IDL (more specifically, we use the \texttt{ch\_pops} procedure).  In practice, the population calculation requires the temperature and electron density, which are outputs from our hydrodynamic calculations. Note that for all our simulations, the ratio $N_2/N_1$ is quite low, with values not exceeding $1$ -- $2\times10^{-5}$ for the young planets, and maximum values well below that for the older planets.\footnote{We also computed the ratio $N_2/N_1$ using the Boltzmann equilibrium equation, and found ratios larger than the ones found by solving the statistical equilibrium equation. The Boltzmann equation predicted larger H$\alpha$ transits, therefore representing an upper limit for the transit calculations. In general, though, these transit depths were not larger by more than a factor of 2 of the ones shown in the present paper.} However, as hydrogen is so abundant, we still predict a reasonably large transit in H$\alpha$, as we will show in Section  \ref{ha_sec}.

\subsection{Ly$\alpha$ transits}\label{lya_sec}
\begin{figure}
\begin{center} 
\includegraphics[scale=0.47]{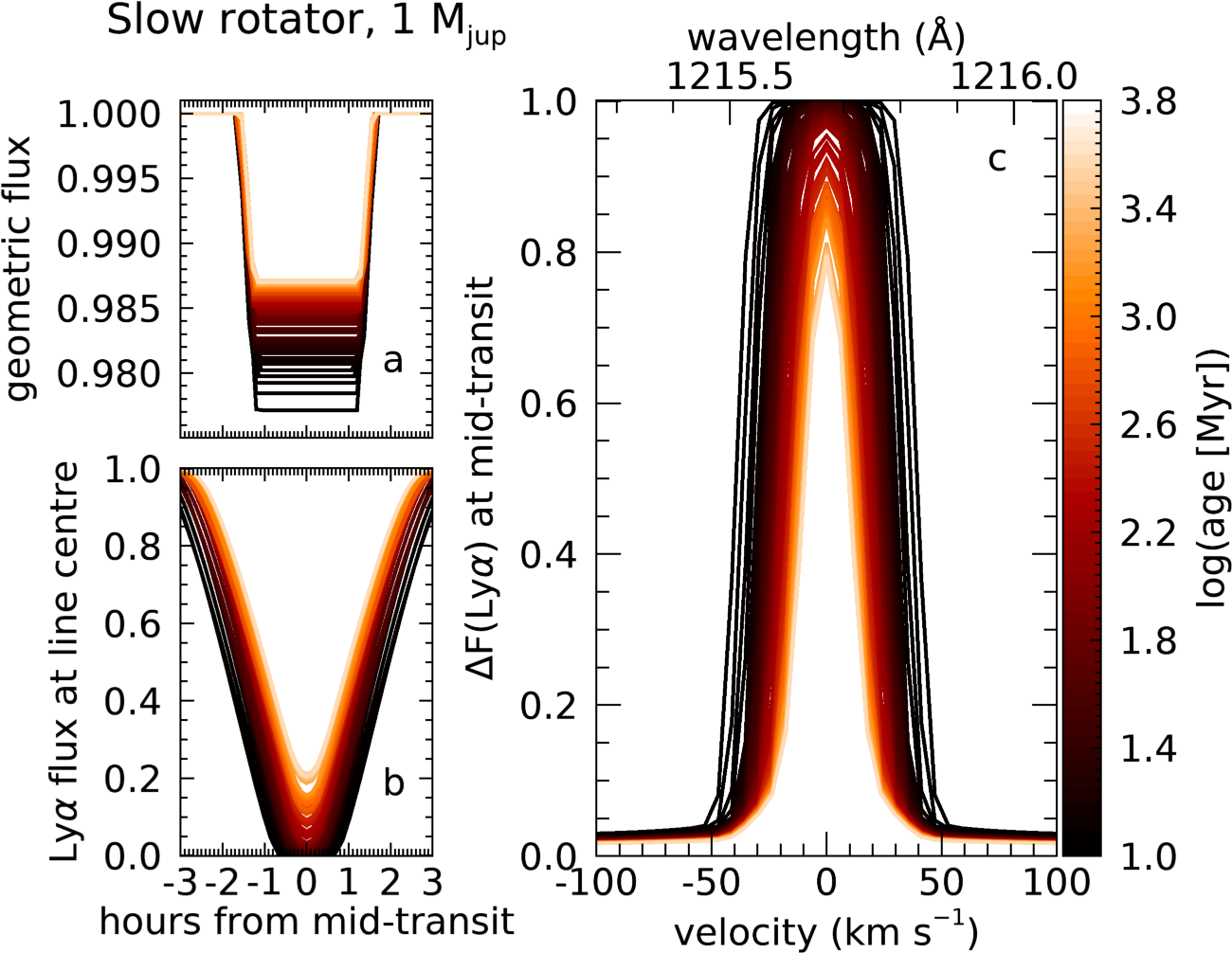}\\ \medskip
\includegraphics[scale=0.47]{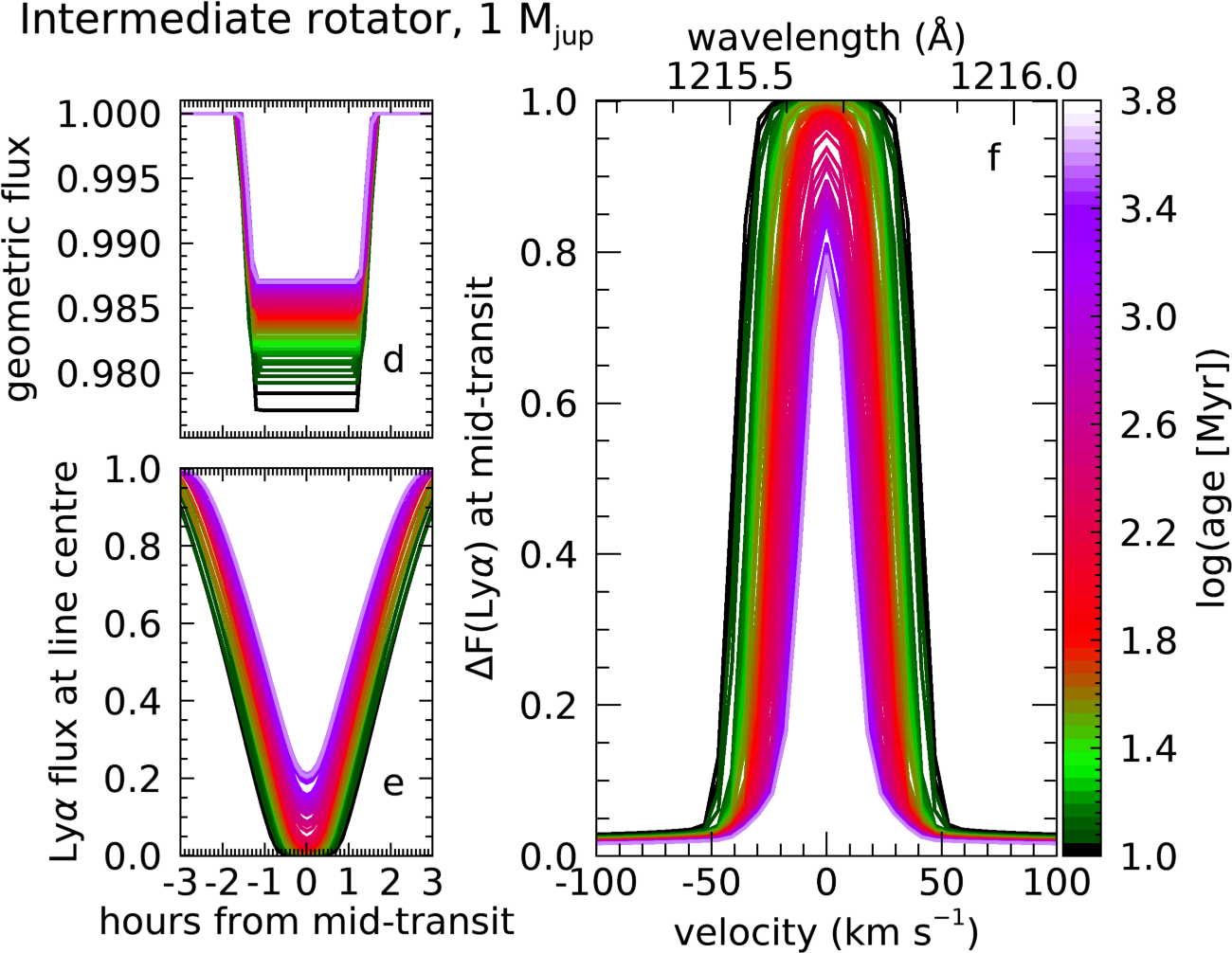}\\ \medskip
\includegraphics[scale=0.47]{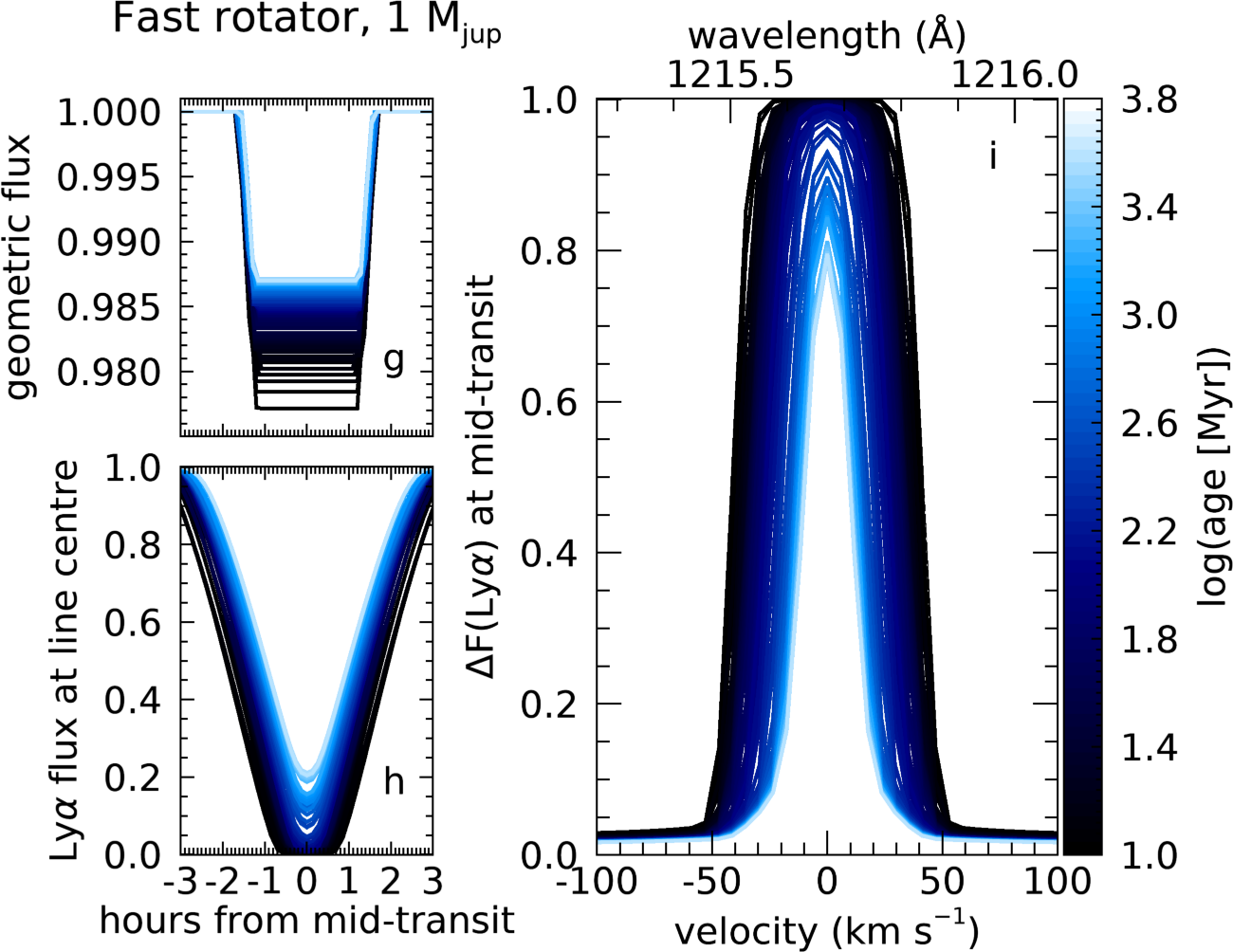}
\caption{Ly$\alpha$ transit simulations for the 1-$M_{\rm jup}$ planet. Results are separated into three blocks related to the host star stellar rotation: slow (a-c), intermediate (d-f) and fast (g-i) rotators. Panels a, d, and g show the lightcurves of the geometric transit. Panels b, e and h show the  lightcurves as observed in Ly$\alpha$ line centre. Panels c, f and i are the mid-transit obscured flux ($\Delta F_\nu$, Equation \ref{eq.deltaF}) of the Ly$\alpha$ line profile, as a function of Doppler velocity and wavelength. Note that $\Delta F_\nu$ does not reach zero and is offset by a couple \% due to the absorption of the  opaque disc of the planet. For each plot,  colour represents age, which are indicated in the colour-bars. }
\label{fig:red_lya}
\end{center}
\end{figure}

Figure \ref{fig:red_lya} shows our predicted transit signatures in the Ly-$\alpha$ line for the 1-$M_{\rm jup}$ planet. For each block of three panels in Figure \ref{fig:red_lya}, colour indicates the planetary age as given by the relevant colour bar. In the three upper plots (a-c), we show the line profiles of hot Jupiters orbiting a slowly rotating host; the central plots (d-f) consider an intermediate rotator, while lower plots (g-i) correspond to a fast rotating host star. The lightcurves resulting from the geometric transit of the planet itself are shown in panels a, d and g. Therefore, these plots simply reveal the transit effect of the evolving planetary radius, where the transit depths can be converted in $(R_{\rm pl}/R_*)^2$. 

Panels b, e and h of Figure \ref{fig:red_lya} display predicted transit lightcurves as observed at the Ly$\alpha$ line centre.  These lightcurves feature considerably deeper transits than those resulting from the geometric transit, as neutral hydrogen in the extended atmospheres absorbs strongly in Ly$\alpha$. In addition, the typical Ly$\alpha$ transit duration is approximately 2 hours longer than the geometric transit, with young planets producing even longer durations.  These plots show the evolution of the atmospheric size: as the atmospheres of younger close-in giants are inflated further than those of older planets, they occult more of the stellar disc, thus resulting in a greater transit duration and depth in Ly$\alpha$. 

As Ly$\alpha$ observations are not possible at the line centre, in panels c, f and i of Figure \ref{fig:red_lya}, we show the predicted  Ly$\alpha$ line profile at mid-transit. In these panels, we show the occulted flux $\Delta F_\nu$ (Equation \ref{eq.deltaF}) as a function of wavelength and Doppler velocity. These profiles show saturation of Ly$\alpha$ absorption at line centre and larger absorption in the line wings  for younger planets. This is a direct result of the stronger atmospheric escape experienced by these planets, resulting in higher velocities of the escaping neutral hydrogen atoms. The symmetric line profiles are a result of our hydrodynamical models being spherically symmetric. A more realistic simulation, including for example interaction with stellar wind and acceleration due to radiation pressure, would present asymmetries in blue and red wings of the line profile \citep[e.g.,][]{2014A&A...562A.116K,2016A&A...591A.121B,2018MNRAS.479.3115V,2018MNRAS.481.5296V,2019ApJ...873...89M, 2019MNRAS.487.5788E}. 

Across the three considered host-star rotations, a  faster rotating host leads to  longer and deeper  Ly$\alpha$ transits (line centre) for a given planetary age below $\sim 800$~Myr. These younger planets also show wider line profiles at mid-transit, when orbiting fast-rotating stars. This is because planets orbiting more active stars receive more $F_{\rm EUV}$, which then creates stronger evaporation.

\begin{figure}
\begin{center} 
\includegraphics[scale=0.47]{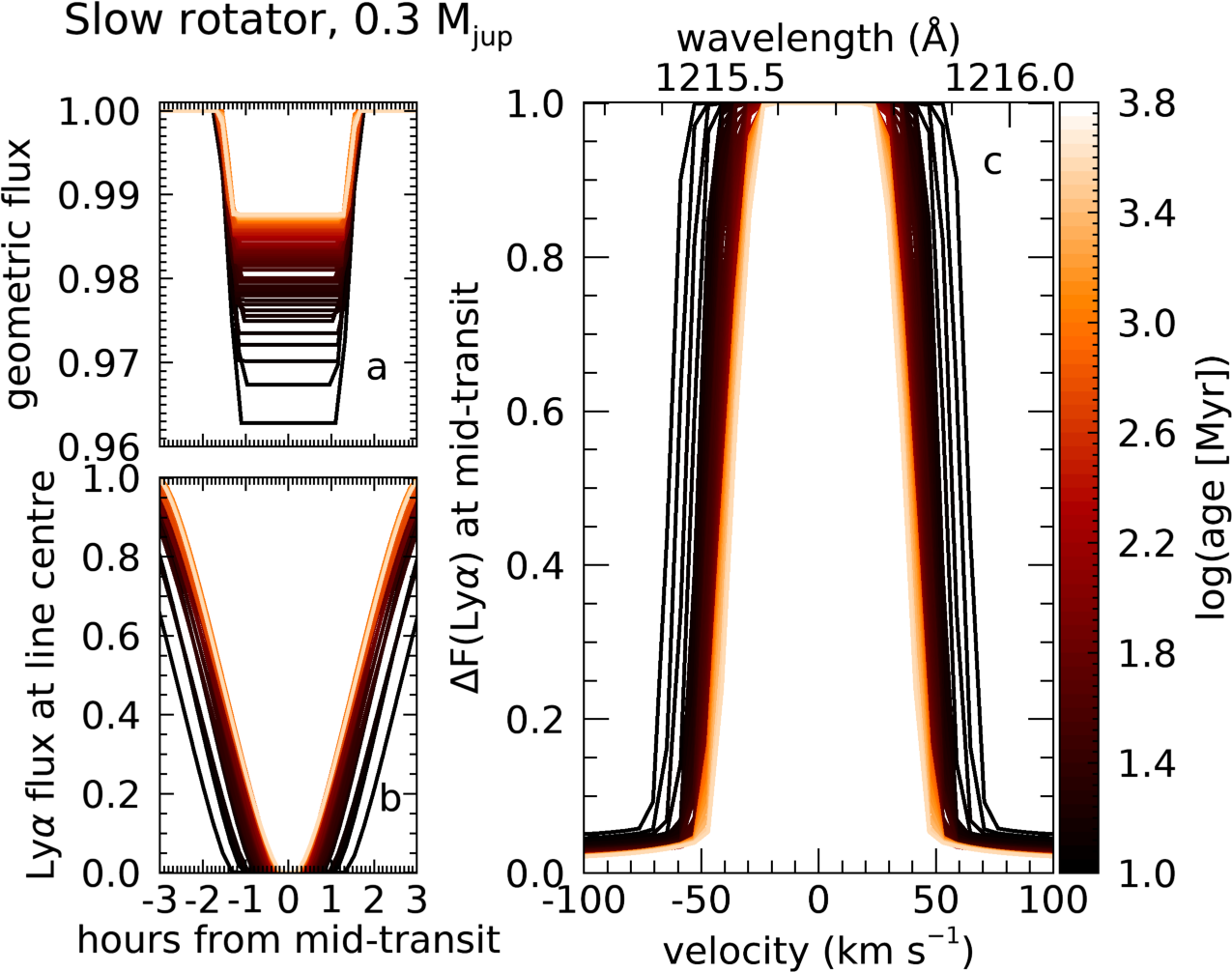}\\ \medskip
\includegraphics[scale=0.47]{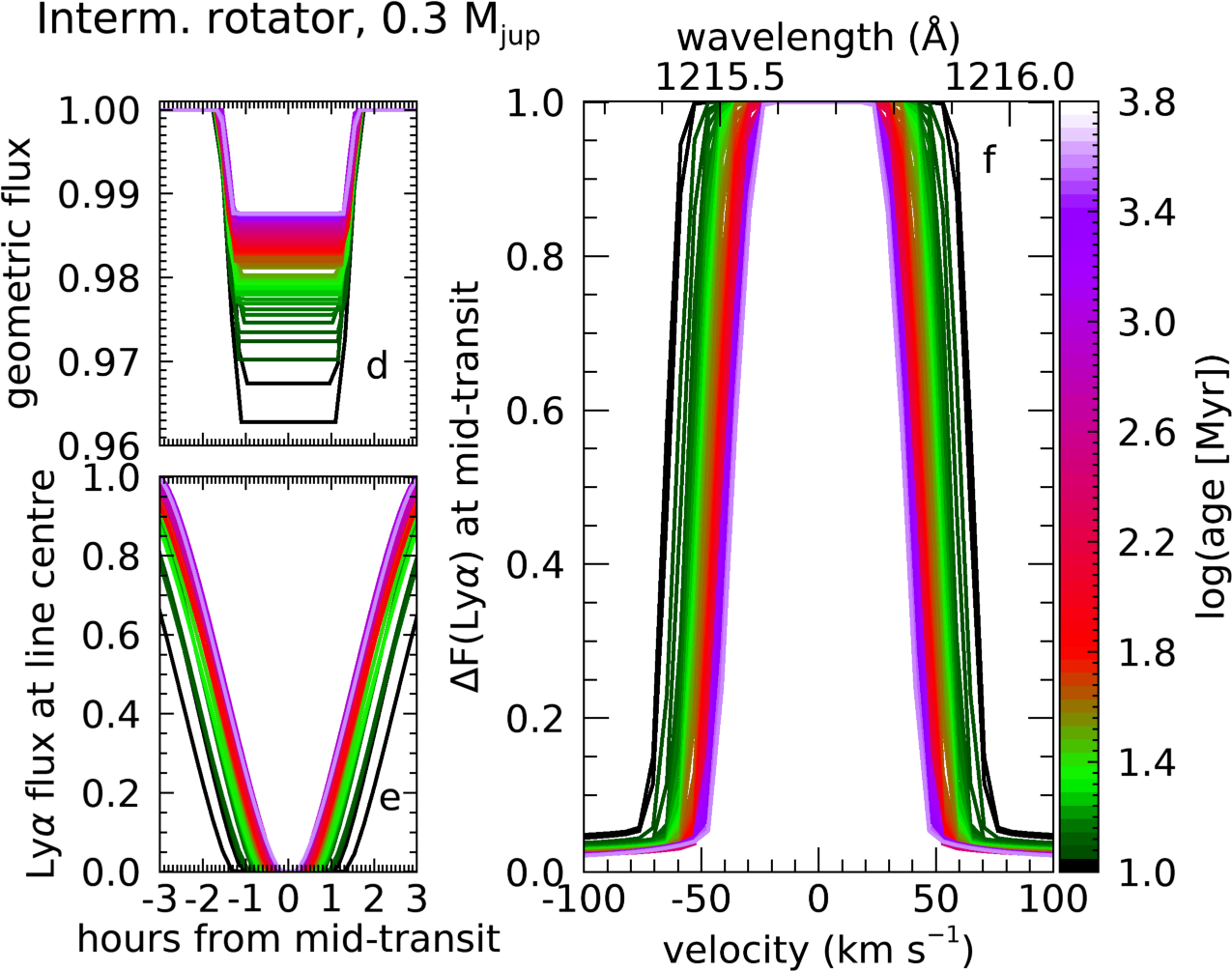}\\ \medskip
\includegraphics[scale=0.47]{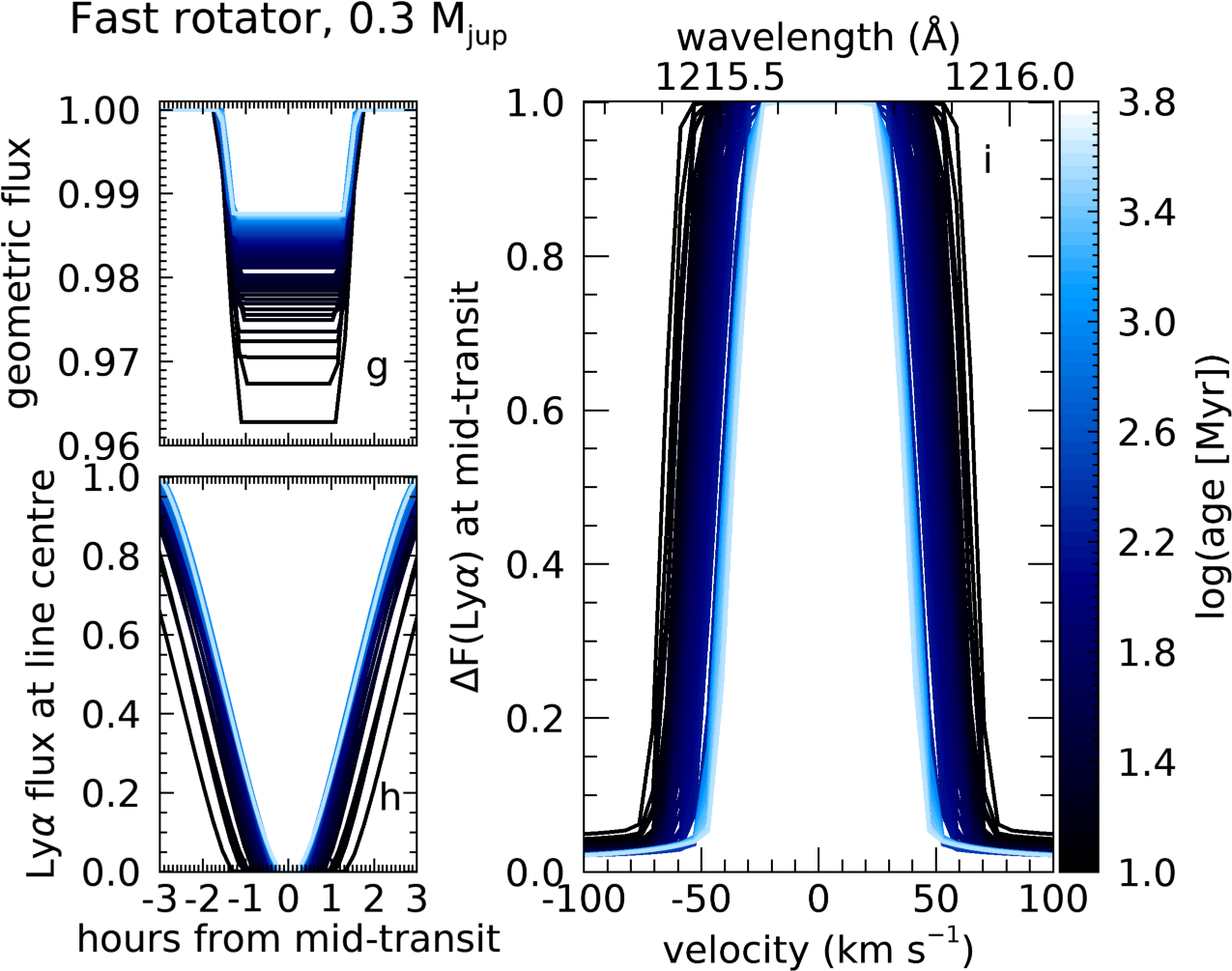}
\caption{Same as in Figure \ref{fig:red_lya}, but for the 0.3-$M_{\rm jup}$ planet.}
\label{fig:red_lya__03MJUP}
\end{center}
\end{figure}

In Figure \ref{fig:red_lya__03MJUP}, we present the same analysis as that shown in Figure \ref{fig:red_lya},  only now we consider the 0.3-$M_{\rm jup}$ planet. It is clear that each of the Ly$\alpha$ transit profiles exhibit more extreme characteristics for this less massive planet: longer duration, deeper transits and wider line profiles. These are due to both a greater inflation of the atmosphere and faster moving neutral hydrogen atoms. 

\subsection{H$\alpha$ transits} \label{ha_sec}
Similar to our predictions of Ly$\alpha$ transits, here we present our estimates of H$\alpha$ transits. An advantage of observing in H$\alpha$ is that the line centre is not contaminated as in the Ly$\alpha$ line and, perhaps, most importantly, they can be conducted with ground-based spectrographs. 

\begin{figure*}
\begin{center} 
\includegraphics[scale=0.47]{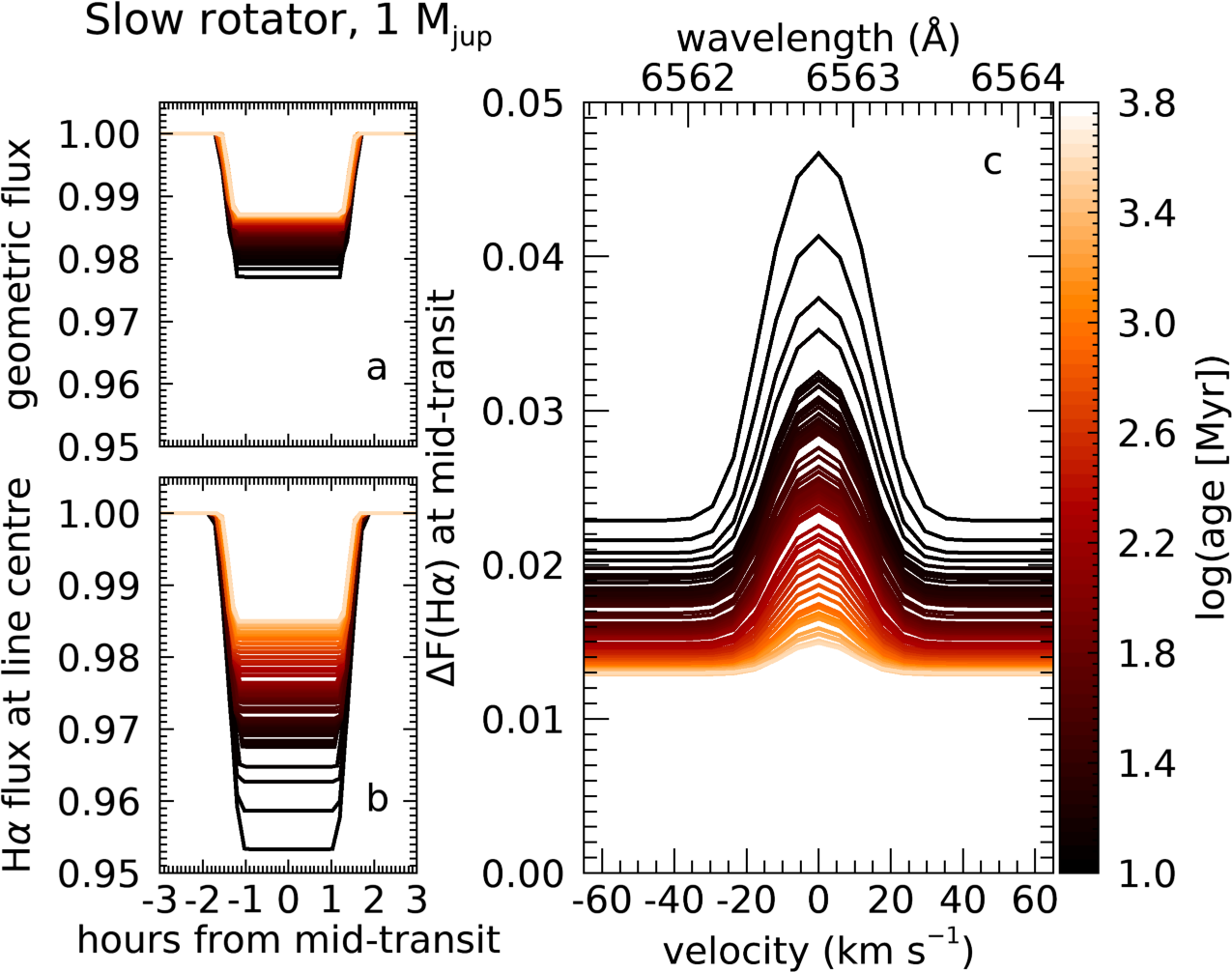} \hspace{0.3cm} \includegraphics[scale=0.47]{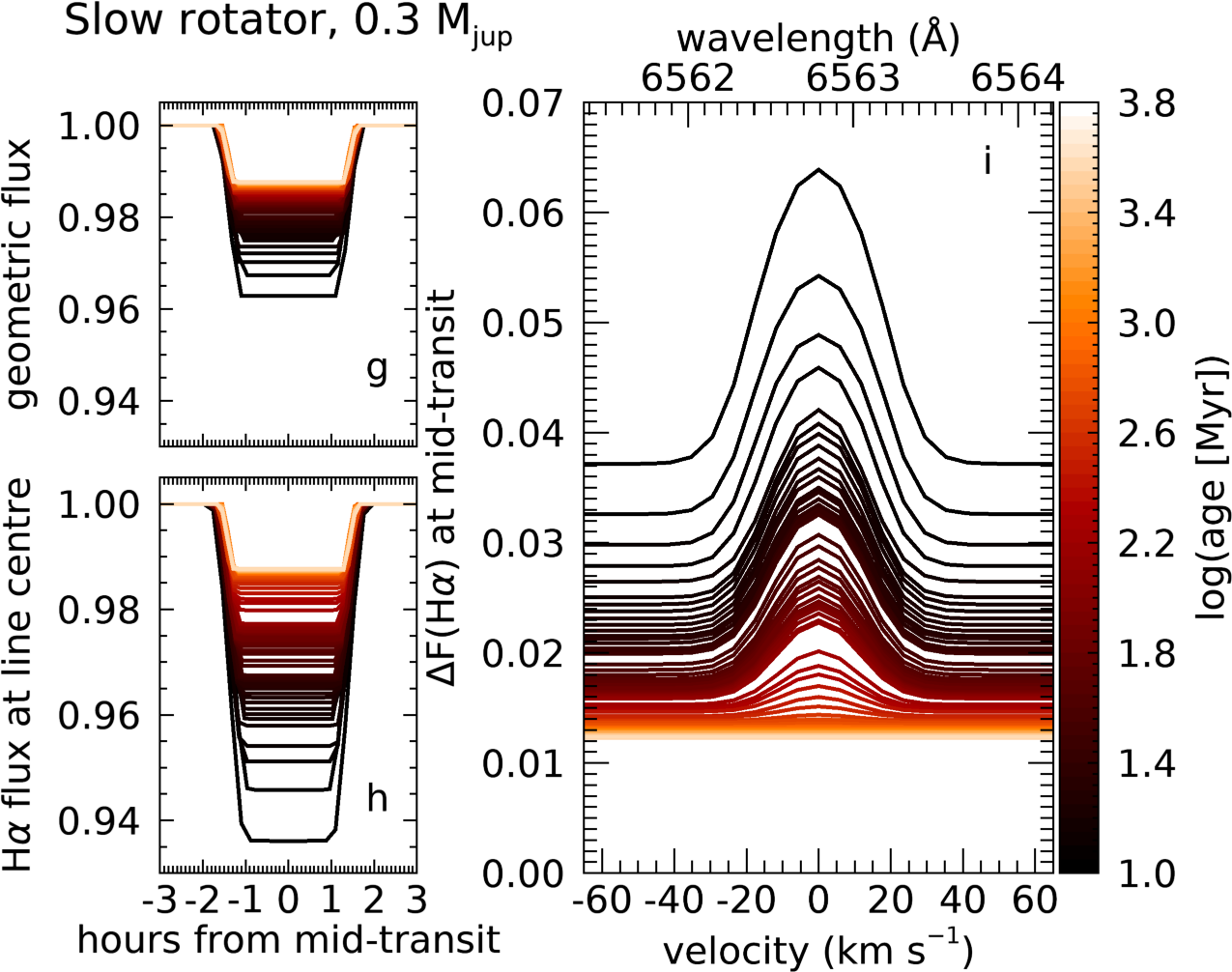}\\  \medskip
\includegraphics[scale=0.47]{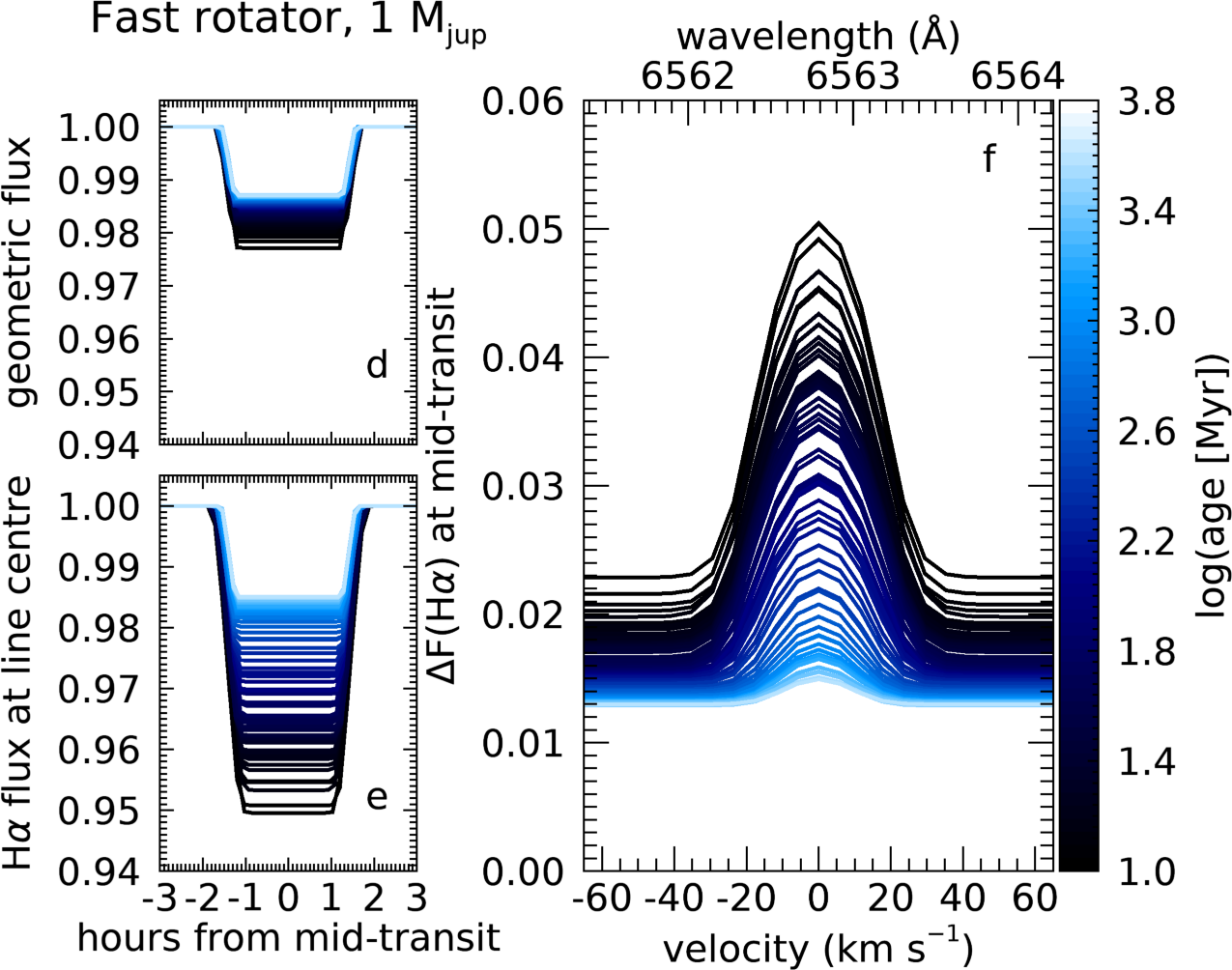} \hspace{0.3cm}  \includegraphics[scale=0.47]{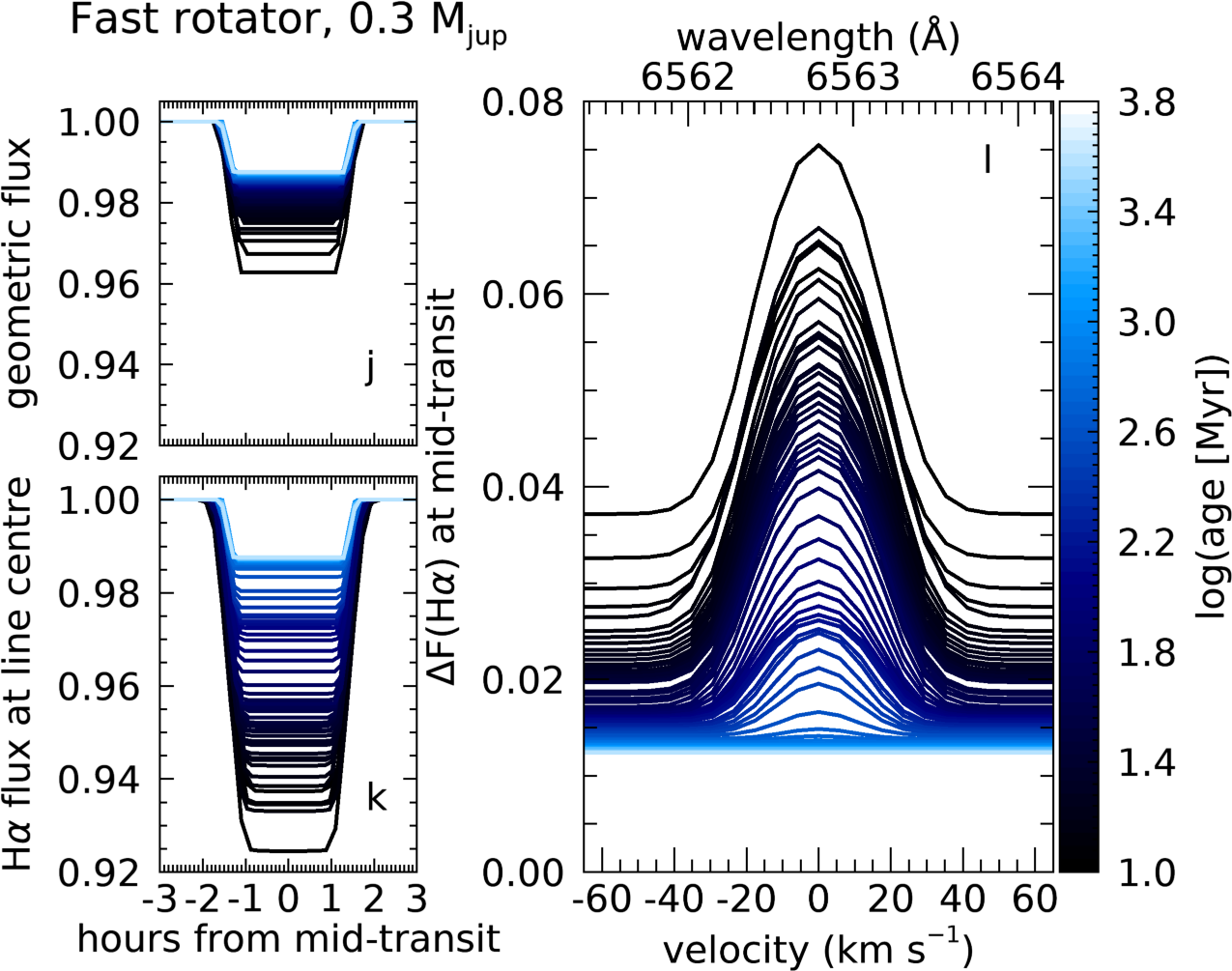}
\caption{Same as in Figure \ref{fig:red_lya}, but for the H$\alpha$ transits of the 1-$M_{\rm jup}$ planet (left) and 0.3-$M_{\rm jup}$ planet (right). The host stars are assumed to start their lives as slow (top), and fast (bottom) rotators. }
\label{fig:red_ha}
\end{center}
\end{figure*}

Figure \ref{fig:red_ha} shows the H$\alpha$ transits for the 1-$M_{\rm jup}$ and 0.3-$M_{\rm jup}$ planets, left and right panels respectively, in a similar design as Figures \ref{fig:red_lya} and \ref{fig:red_lya__03MJUP}. We omit here the intermediate rotator for brevity and only show results for the slow (top panels) and fast (bottom) host-star rotations. The trends in the H$\alpha$ transits are {\it qualitatively} similar to the ones seen in the Ly$\alpha$ ones: the H$\alpha$ transits exhibit greater depths and duration than the geometric lightcurves due to absorption of H$\alpha$ photons by neutral hydrogen in the first excited state in the evaporating atmosphere. Likewise, we find that younger planets produce deeper, longer and broader H$\alpha$ transits compared to their older counterparts. For a given age below $\sim 800$~Myr ($\log ({\rm age})=2.9)$, the obscuration is larger for planets orbiting fast rotators, due to the higher $F_{\rm EUV}$ of these stars. However, Ly$\alpha$ and H$\alpha$ transits are  {\it quantitatively} different.

\begin{figure}
\begin{center} 
\includegraphics[width=0.47\textwidth]{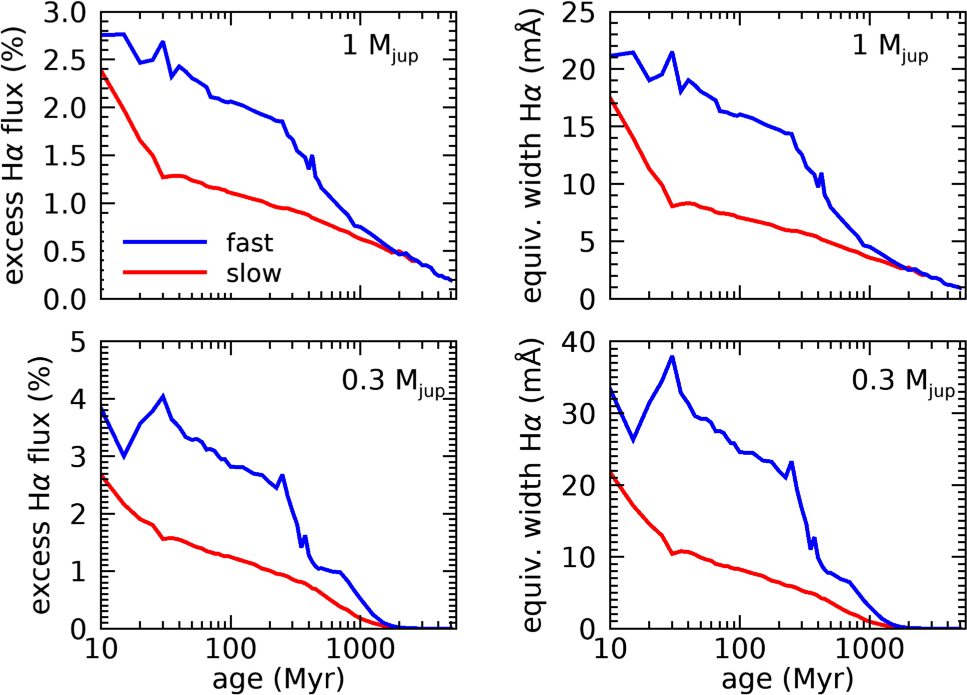}
\caption{Left: Evolution of transit depths, in excess of the geometric transit, in the H$\alpha$ line at line centre and at mid-transit. Right:  Evolution of the mid-transit equivalent width (Equation \ref{eq.width}), integrated over $\pm 500$~km/s. Top panels are for the 1-$M_{\rm jup}$ planet simulations while the bottom panels are for the 0.3-$M_{\rm jup}$ case. Red and blue curves represent the rotation of the host star, slow and fast, respectively.}
\label{fig:Halpha_transit_flux}
\end{center}
\end{figure}

Contrary to the saturation often seen in Ly$\alpha$ for several ages and host star rotations, H$\alpha$ absorption is considerably weaker. This is due to the significantly lower density of hydrogen  in the first excited state $n=2$ compared to the ground state $n=1$. As a result, the transit depths we predict in our models reach, at line centre, at most $\Delta F_\nu  \sim$5 -- 8\% (a $\sim$ 3 -- 4\% excess from the geometric transit) at very young ages reducing, at older ages, to fractions of a percent above the geometric transit, for the 1-$M_{\rm jup}$ planet, or no excess absorption at all for the 0.3-$M_{\rm jup}$ planet. The left panels in Figure \ref{fig:Halpha_transit_flux} show the evolution of the H$\alpha$ excess flux, i.e., above the geometric transit, for the planets simulated here.

Our results, therefore, indicate that H$\alpha$ transits could be more easily detected in planets orbiting young and/or active stars. An extended H$\alpha$ transit has indeed been observed for HD189733b, which orbits a relatively active star  \citep{2012ApJ...751...86J}. These authors report a planet size in H$\alpha$ that is $\sim 1.4 R_{\rm pl}$, or $\Delta F_\nu \simeq 4.7\%$ at line centre at mid-transit. These values are actually not too dissimilar to our models of the  1-$M_{\rm jup}$ planet at younger ages. H$\alpha$ was also detected in the atmosphere of KELT-9b \citep{2018NatAs...2..714Y, 2019AJ....157...69C}.  Although the host is not an active star, it has a strong EUV flux due to its early spectral type. As a result, KELT-9b shows a wide H$\alpha$ line profile ($\sim$ 50 km/s) and a transit depth of 1.8\% (1.15\% in excess of the geometric transit of 0.68\%, \citealt{2018NatAs...2..714Y}). 
Conversely, an H$\alpha$ extended transit was not detected in the less active  KELT-3 system \citep{2017AJ....153..217C}. 

Our models predict that lower gravity giants, like our 0.3-$M_{\rm jup}$ case, would show a larger H$\alpha$ obscuration  than the more massive planets at young ages (compare top and bottom panels in Figure \ref{fig:Halpha_transit_flux}).  However, at older ages, the situation is reversed, with the  0.3-$M_{\rm jup}$ showing no excess H$\alpha$ flux above $\sim$1.2 Gyr.  This is because the extended atmosphere of neutral hydrogen becomes almost entirely in the ground state at older ages. Our result agrees with the lack of absorption observed in the  H$\alpha$ transits of GJ436b \citep{2017AJ....153..217C}, in spite of its impressively long and extended transit in Ly$\alpha$ \citep{2014ApJ...786..132K, 2015Natur.522..459E,2017A&A...605L...7L}. This situation can occur when temperatures are not high enough to excite hydrogen. Indeed, GJ436b has a relatively small EUV flux ($F_{\rm EUV} = 10^3$ erg/cm$^2$/s, \citealt{2015Natur.522..459E}). In our simulations, this flux is reached at an age of $\sim$ 3Gyr, when we already see a lack in H$\alpha$ absorption.

We also calculate the equivalent width of our H$\alpha$ transmission spectra as
\begin{equation}\label{eq.width}
    W_{{\rm H}\alpha}=\int_{v_i}^{v_f} (\Delta F_{\nu} -\Delta F_{\rm geom}) d{v} , 
\end{equation}
where ${v_i}$ and ${v_f}$ represent the velocity range over which the integration is performed and  $\Delta F_{\rm geom}$ is the geometric transit depth ($R_{\rm pl}^2/R_*^2$), which is wavelength-independent. Equation (\ref{eq.width}) is thus the integral of the `excess' transit over a velocity (or wavelength) range. We integrate over our entire simulated velocity range of $\pm 500$ km/s or, equivalently, $\pm 10.9$\AA.  However, we note that decreasing the range of integration to $\pm 50$ km/s  ($\pm 1$\AA) gives the same equivalent width -- this is because the bulk of our predicted absorption occurs within $\pm 50$ km/s. 

The right panels of Figure \ref{fig:Halpha_transit_flux} show the evolution of $W_{{\rm H}\alpha}$ for the  1- (top) and 0.3-$M_{\rm jup}$ (bottom) planets, respectively. Equivalent widths are larger at younger ages for both mass planets, reaching 20 -- 40 m\AA\ for planets orbiting active, young stars. Our values are lower than, but still similar to, the $\sim 65$ m\AA\ observed for the hot Jupiter HD189733b \citep{2018AJ....156..154J}. 

\begin{figure}
\begin{center} 
\includegraphics[width=0.47\textwidth]{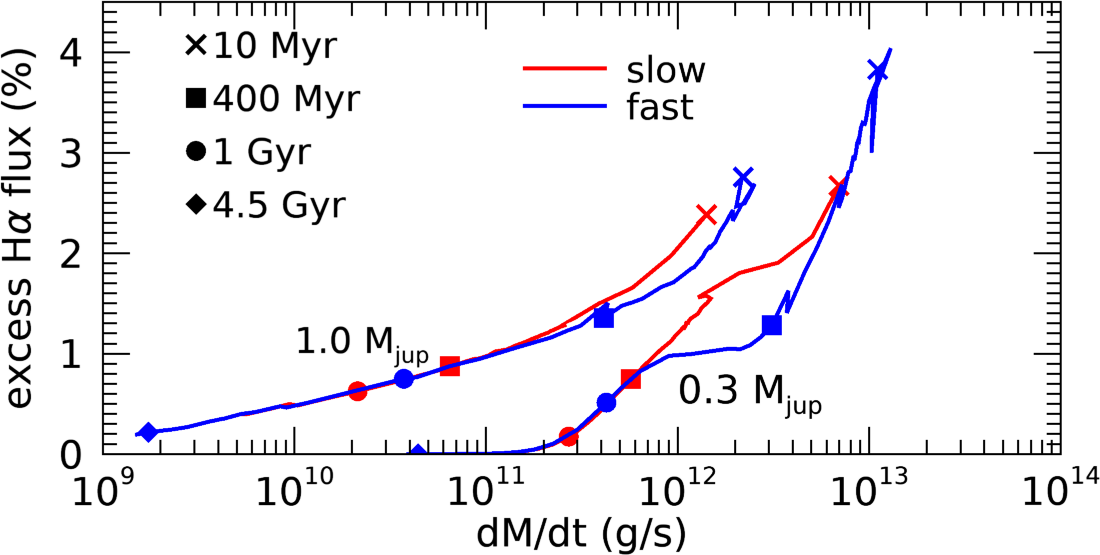}
\caption{Transit depths, in excess of the geometric transit, in the H$\alpha$ line at line centre and at mid-transit versus the predicted escape rate. Symbols mark different evolutionary stage, with the younger systems at the top right of the curves and evolving as going towards the bottom left. The left (right) most curves are for the 1-$M_{\rm jup}$ (0.3-$M_{\rm jup}$) planet simulations.}
\label{fig.excess_mdot}
\end{center}
\end{figure}

Both the excess transit and the equivalent width of the H$\alpha$ line (Figure \ref{fig:Halpha_transit_flux}) show trends that are similar to the $F_{\rm EUV}$ evolution with rotation of the host (Figure \ref{fig.fEUVandrad} top): the curve for fast rotator lies above the curve for slow rotator. However, contrary to the saturation of flux seen at early ages, which is also reflected in the curves of $\dot{M}$ (Figure \ref{fig.evol}), the H$\alpha$ line shows neither a plateau-like saturation in transit depth nor in the equivalent width (Figure \ref{fig:Halpha_transit_flux}). This can be seen in Figure \ref{fig.excess_mdot}, were we plot the excess transit versus the escape rates. At younger ages, from 10 to 400 Myr (between crosses and squares), we see relatively small variations in $\dot{M}$ for planets around fast rotating stars (in the saturated regime), but a significant drop in excess absorption. 

\section{Discussion and conclusions}\label{CONCLUSIONS}
In this work, we studied the evolution of atmospheric escape in close-in giants. For that, we modelled escape of a higher gravity planet (1.0 $M_{\rm jup}$) and a lower gravity one (0.3 $M_{\rm jup}$). Both planets were assumed to be at 0.045~au, orbiting  stars whose EUV fluxes evolve from early ages (10 Myr) to about 5 Gyr. Given that the host star's fluxes evolve differently depending on their initial rotation, we considered three evolutionary paths for the EUV flux, by following host stars that initiated their lives as slow, intermediate or fast rotators \citep{Johnstone2015a}. In our models, we also took into  account contraction of the planet during its evolution \citep{Fortney2010}. 

To probe the entire evolution, for the two different planet masses, and three different stellar rotations, we run more than 300 simulations of atmospheric escape, using a hydrodynamic escape model.  We found that younger planets experience higher levels of atmospheric escape, owing to a favourable combination of higher irradiation levels and weaker planetary gravity (due to their larger radii). Lesser mass planets  experience even higher escape rates than the more massive ones. Altogether, the lower gravitational potential of the less massive planet results in stronger planetary winds, with higher velocities and mass-loss rates. 

In spite of the relatively large escape rates ($10^9$ -- $10^{12}$ g/s), we estimated that, over 5 Gyr,  the 1-$M_{\rm jup}$ planet would lose at most 1\% of its initial mass due to escape. However, for the 0.3-$M_{\rm jup}$ planet ($10^{10}$ -- $10^{13}$ g/s), up to 20\% of its mass could evaporate through hydrodynamic escape. This strong escape in the lower mass planet is in line with observations of the short-period `Neptunian desert', which shows a lack of planets at $\lesssim 5$--$10$-day orbit, with masses between 0.03 and 0.3 $M_{\rm jup}$ \citep{2016A&A...589A..75M}. The desert is suggestive of significant mass loss in planets with masses $\lesssim 0.3 ~M_{\rm jup}$, which is indeed confirmed by our models. 
 
Using the results of our hydrodynamic models, we then computed spectroscopic transits in  both Ly$\alpha$ and H$\alpha$ lines.  The Ly$\alpha$ line is saturated at line centre and at mid-transit (we assume no impact parameter for simplicity), while H$\alpha$ transits produce transit depths of at most 3 -- 4\% in excess of their geometric transit at younger ages. Planets older than 200 Myr (or 600 Myr, if orbiting around fast rotating star) would have H$\alpha$ absorption  $\lesssim$1\% in excess of the geometric transit. 
While at older ages ($\gtrsim$1.2 Gyr), Ly$\alpha$ absorption is still significant (and the line could even be saturated, as in the case of the lower mass planet), the H$\alpha$ absorption nearly disappears. This is because the extended atmosphere of neutral hydrogen becomes almost entirely in the ground state at such ages. 

Our models, however, predict line profiles that are symmetric, in contrast to asymmetric profiles derived from observations (e.g., HD189733b (\citealt{Etangs2012a}); HD209458b (\citealt{2003Natur.422..143V, Ben-Jaffel2008}); GJ436b \citep{2015Natur.522..459E}). This is because our escape models are spherically symmetric. To overcome this limitation, we would need to consider additional physical ingredients in our models, such as charge-exchange between stellar wind protons and atmospheric neutrals, acceleration of neutral atoms due to radiation pressure and interaction of planetary flows with a stellar wind. Several multi-dimensional simulation studies have moved into this direction \citep[e.g.,][]{2014A&A...562A.116K,2015ApJ...813...50K,2016A&A...591A.121B,2018MNRAS.479.3115V,2018MNRAS.481.5296V,2019ApJ...873...89M, 2019MNRAS.487.5788E,2019MNRAS.483.2600D,2019arXiv190600075D, carolan2019}. These physical mechanisms are also required to reproduce the large velocities seen in Ly$\alpha$. While these model provide more physical intuition in the details of the atmospheric escape, they can be computationally expensive.

The advantage of our models, versus more complex ones, is thus the ability of computing a large number of simulations due to their low computational cost. As a result, we can make general predictions that could help guide observations. We showed here that younger systems show deeper transits as well as larger widths in both H$\alpha$ and Ly$\alpha$ lines. In theory, this  indicates that  atmospheric escape is easier to detect during early ages. However, in practice, at younger ages, stars are more active, generating  variability that could mimic signatures of atmospheric transits \citep{2016MNRAS.462.1012B,2018AJ....156..154J}. This effect makes escape observations at  young ages and/or around active stars more difficult to interpret and, thus, models, like ours, can assist in the interpretations of transit signatures. 

\section*{Acknowledgements}
We acknowledge funding from the Irish Research Council Consolidator Laureate Award 2018. CHIANTI is a collaborative project involving George Mason University, the University of Michigan (USA), University of Cambridge (UK) and NASA Goddard Space Flight Center (USA). The authors thank C.~Villarreal D'Angelo and A.~Esquivel for their insightful comments and discussion.  


\appendix
\section{Analytical fits to mass-loss rates and terminal velocities} \label{evol_analt}
Here, we present power-law fits of the evolutionary curves of mass-loss rate and terminal velocity presented in Figure \ref{fig.evol}. These quantities take the form
\begin{equation}\label{eq:analt1}
    \frac{\dot{M}}{[\rm g/s]}= 10^a     \left(\frac{\rm age}{\rm [Myr]}\right)^{-b} 
    \end{equation}
\begin{equation}\label{eq:analt2}
        \frac{u_{ \rm term}}{[\rm km/s]}=  10^c \left(\frac{\rm age}{\rm [Myr]}\right)^{-d} 
  \end{equation}
where $a$, $b$, $c$ and $d$ are shown in Tables \ref{tab:table_analt} and \ref{tab:table_analt2}. These expressions give errors of up to 1.5\% for $\dot{M}$ and 1.6\% for u$_{\rm term}$. 
  
\begin{table}
	\centering
	\caption{Coefficients to fits given in Equations (\ref{eq:analt1}) and (\ref{eq:analt2}) for the three considered host star rotations at various ages, as shown in Figure \ref{fig.evol}, for the 1-$M_{\rm jup}$  planet.}
	\label{tab:table_analt}
	\begin{tabular}{lccccc} 
		\hline
		age range [Myr] & $a$  & $b$ & $c$  & $d$\\
		\hline 
		Slow  & & & &  \\
        10 -- 30 &   13.8  &    1.58   &   1.71  &     0.107  		\\
        30 -- 250  &   12.2  &    0.532 &   1.66  &    0.0902  \\
        250 -- 5000  &    14.4  &    1.37   &   1.79  &    0.175  \\
		Intermediate & & & &  \\
		10 -- 30 &    13.1  &    0.714   &    1.78  &    0.128  \\
        30 -- 250  &   12.9  &     0.621   &    1.71  &     0.0991 \\
        250 -- 5000  &    15.5  &    1.69   &   1.75  &    0.116  \\
		Fast  & & & &  \\
        10 -- 30 &   12.2  &     -0.0918   &   1.94  &    0.174   \\
        30 -- 250  &   12.7  &     0.287  &   1.70  &    0.0695   \\
        250 -- 5000  &   17.4  &    2.24   &    1.67  &     0.0414  	\\
        	\hline			\end{tabular}
\end{table}

\begin{table}
	\centering
	\caption{The same as in Table \ref{tab:table_analt}, but for the 0.3-$M_{\rm jup}$ planet.}
	\label{tab:table_analt2}
	\begin{tabular}{lccccc} 
		\hline
		age range [Myr] & $a$  & $b$ & $c$  & $d$\\
		\hline
		Slow  & & & &  \\
        10 -- 30 &   14.4  &     1.52   &   1.91  &     0.104   		\\	
        30 -- 250  &   12.7  &    0.342   &   1.82  &     0.0758  	\\
        250 -- 5000  &    14.3  &    0.977 &    1.95  &     0.160 	\\
        Intermediate & & & &  \\
        10 -- 30 &   13.7  &     0.630  &   1.91  &     0.106  	\\
        30 -- 250  &    13.4  &     0.488   &    1.86  &     0.0897  	\\
        250 -- 5000  &    15.3  &     1.26   &   1.93  &    0.125  	\\
        Fast  & & & &  \\
        10 -- 30 &   12.9  &     -0.138   &   2.02  &     0.136  	\\
        30 -- 250  &    13.4  &     0.227   &    1.85  &    0.0662   	\\
        250 -- 5000  &   17.0  &     1.76   &   1.87  &    0.0716  	\\
		\hline
			\end{tabular}
\end{table}

 Equation \ref{eq:analt1} offers an alternative to the widely used energy-limited approximation for $\dot{M}$. In addition, a priori knowledge of the effective radius $R_{\tau=1}$ included in the energy limited approximation (Equation \ref{eq: E_lim}) is not required by our expression for mass-loss rates.

\bsp	
\label{lastpage}
\end{document}